\documentclass[journal]{IEEEtran}
\usepackage{cite}

\ifCLASSINFOpdf
\usepackage[pdftex]{graphicx}
\DeclareGraphicsExtensions{.pdf,.jpeg,.png}
\else
\usepackage[dvips]{graphicx}
\DeclareGraphicsExtensions{.eps}
\fi
\usepackage{epsfig,epsf,epstopdf,graphicx}
\usepackage[cmex10]{amsmath}
\usepackage{amssymb}
\usepackage{amsthm }
\usepackage{nicefrac}
\usepackage{hyperref}
\usepackage{xcolor}
\hypersetup{colorlinks=true}
\usepackage{tabularx}

\theoremstyle{theorem}

\theoremstyle{definition}

\theoremstyle{plain}

\theoremstyle{plain}

\newcommand{\sbb}{{\textbf{s}}}

\newcommand{\zb}{{\textbf{z}}}

\newcommand{\nb}{{\textbf{n}}}

\newcommand{\rb}{{\textbf{r}}}

\usepackage{algorithm}
\usepackage{multirow}
\usepackage{algcompatible}
\usepackage[bottom]{footmisc}

\PassOptionsToPackage{dvipsnames}{xcolor}
\usepackage{tikz}
\usetikzlibrary{calc}

\ifCLASSOPTIONcompsoc
  \usepackage[caption=false,font=normalsize,labelfont=sf,textfont=sf]{subfig}
\else
  \usepackage[caption=false,font=footnotesize]{subfig}
\fi
\usepackage{fixltx2e}
\usepackage{afterpage}
\usepackage{float}
\usepackage{stfloats}

\makeatletter

\begin{document}

%
\title{Markovian Block Sparse Signal Detection Using One Bit Measurements}

\author{Alireza Hariri, \IEEEmembership{Member,~IEEE,}
Hadi~Zayyani, \IEEEmembership{Member,~IEEE,} and
Mehdi Korki, \IEEEmembership{Member,~IEEE}

\thanks{A.~Hariri, is with the Department
of Electrical and Computer Engineering, Qom University of Technology (QUT), Qom, Iran (e-mails: alireza.hariri1@gmail.com).}
\thanks{H.~Zayyani, is with the Department
of Electrical and Computer Engineering, Qom University of Technology (QUT), Qom, Iran (e-mails: zayyani@qut.ac.ir).}
\thanks{M.~Korki is with the School
of Science, Computing, and Engineering Technologies, Swinburne University of Technology, Melbourne, Australia (e-mail: mkorki@swin.edu.au).}
\vspace{-0.5cm}}


\maketitle
\thispagestyle{plain}
\pagestyle{plain}

\begin{abstract}
This paper presents a novel sparse signal detection scheme designed for a correlated Markovian Bernoulli-Gaussian sparse signal model, which can equivalently be viewed as a block sparse signal model. Despite the inherent complexity of the model, our approach yields a closed-form detection criterion. Theoretical analyses of the proposed detector are provided, including the computation of false alarm probability and detection probability through closed-form formulas. Simulation results compellingly demonstrate the advantages of our proposed detector compared to an existing detector in the literature.
\end{abstract}


%
\IEEEpeerreviewmaketitle

\section{Introduction}
\label{sec:Intro}
\IEEEPARstart{T}{he} detection of a signal in noise is a fundamental signal processing task which has wide applications in radar, spectrum sensing, wireless sensor networks, and etc \cite{Kaydet98}. In many situations, the signal is sparse which means that most of the signal samples (or signal samples in other domain) are zero or near zero and a few of signal samples are nonzero. Hence, sparse signal detection is a topic in detection theory which is the main subject of this paper. The sparse signal detection dates back to pioneering works \cite{Duarte06}, \cite{Haupt08} which discussed the problem of detection of an sparse signal in noise. In \cite{Duarte06}, they showed how the compressed sensing principles can solve the detection of sparse signals without the need to recover the thorough signal. \cite{Haupt08} proposed a multi-step adaptive resampling procedure for detection of a high-dimensional sparse signal in noise. In \cite{Wima17}, a partial support set estimation method is used for the problem of sparse signal detection from compressive measurements. \cite{Choi16} proposed a sparse signal detection scheme of sparse shift keying signal using a Monte-Carlo Expectation-Maximization (EM) algorithm. Moreover, there is a work that discusses the problem of detection of a general non-sparse signal from compressive measurements in the Neyman-Pearson framework \cite{Naga18}. In addition, in \cite{Kafle18}, a Bayesian framework is suggested to solve the problem of sparse signal detection using a Laplace prior for modeling the sparsity. Besides, \cite{Wang18} proposed a detection rule of a sparse signal in a sensor network using locally most powerful tests. In sequel, in \cite{Li20}, the problem of detection of sparse stochastic signal in a battery-powered sensors network is investigated using the locally most powerful tests in which there is censoring sensors to yield energy efficiency of the network. In addition, after that, the generalized locally most powerful test is used for the same problem \cite{Mohammadi22}. Also, a detection algorithm combined with Orthogonal Matching Pursuit (OMP) and maximum likelihood (ML) is also presented especially for space shift keying scenario \cite{Zuo21}. Moreover, \cite{Han22} proposed a two stage approach to detect sparse signals from compressive measurements. Furthermore, two secure distributed detection of sparse signals are developed in the literatures \cite{Li19}, \cite{Li20secure}. In \cite{Li19}, the problem of detection of a sparse signal based on falsified compressive measurements in presence of an eavesdropper is investigated, while the same problem using a falsified censoring strategy is discussed in \cite{Li20secure}.

There are also some sparse signal detection schemes with quantized measurements to alleviate the communication burden of the scheme \cite{Wang19_1}, \cite{Wang19_2}, \cite{Quan24}. In \cite{Wang19_1}, a quantized local most powerful detector is proposed and the quantizer's thresholds are designed to have a near optimal detection performance. \cite{Wang19_2} discusses the same approach with the difference that both the noise and dominant elements in sparse signal follows a Generalized Gaussian Distribution (GGD). The authors in \cite{Quan24} investigated the distributed detection of sparse stochastic signals with quantized measurements under Byzantine attacks using the Bernoulli-Gaussian (BG) distribution to model sparse signals. In the limiting case of one bit quantization, some works on signal detection are reported in the literature. The detection of signals with one bit measurements and investigating the degradation of detection is discussed in an early pioneering paper in 1995 \cite{Willett95}. In \cite{Fang13}, a generalized likelihood ratio test is used for one bit deterministic signal detection. Moreover, a double detector scheme is proposed in \cite{Zayy16} to solve the problem of sparse signal detection using one bit compressed sensing measurements. In addition, one bit local most powerful test is suggested to solve the weak signal detection from one bit measurements under observation model uncertainties \cite{Wang19_3}. In a different way, \cite{Li19_1} proposes a sparse signal detection scheme from one bit local likelihood ratios. Also, the problem of sparse signal detection in the tree structured sensors network is discussed in \cite{Li20tree}. Furthermore, there are some works of one bit signal detection in the application of spectrum sensing \cite{Ali18}, \cite{Ali19}, \cite{Zayy20}, \cite{Zhao21}. In \cite{Ali18}, a cooperative wideband sensing based on fast Fourier transform-based one bit quantization sensing is used. Also, in \cite{Ali19}, an ultra low power wideband sensing is proposed which uses one bit Analog to Digital Converters (ADC). \cite{Zayy20} investigates the problem of one bit spectrum sensing in cognitive radio sensors networks using a simple correlated signal model. In \cite{Zhao21}, an approach based on the eigenvalue moment ratio is used for one bit spectrum sensing. Recently, \cite{Ni23} proposes a simultaneously detecting and localizing of a signal with one bit ADC in which an infinite bit Generalized Likelihood Ratio Test (GLRT) is used for detection.

The aforementioned works, use different sparse signal models (priors) such as Bernoulli-Gaussian (BG) \cite{Wang18}, \cite{Li20}, \cite{Wang19_1}, \cite{Li19}, \cite{Li19_1}, Laplace \cite{Kafle18}, and GGD \cite{Wang19_2}. Also, they used different detection strategies such as Likelihood Ratio Test (LRT) \cite{Zayy20}, \cite{Fang13}, \cite{Zayy16}, Local Most Powerful Detector (LMPT) \cite{Wang18}, \cite{Li20}, and Generalized LMPT (GLMPT) \cite{Mohammadi22}. Moreover, some of them use directly the sign of the signal \cite{Ali18},\cite{Ali19}, \cite{Zhao21}, \cite{Zayy20}, while others use one bit compressive measurements \cite{Zayy16}, \cite{Wang18}, \cite{Wang19_1}, \cite{Li19}, \cite{Li19_1}. While very few papers have addressed the correlation between successive samples \cite{Zayy20}, to the best of our knowledge, there is currently no existing research that delves into the consideration of a correlated sparse signal model or a block sparse signal model.

This paper employs a Markovian Bernoulli-Gaussian (BG) sparse signal model \cite{Habibi21} capable of effectively capturing block sparse signals. We address the sparse signal detection problem utilizing one-bit measurements, specifically focusing on the sign of the signal. We derive the Likelihood Ratio Test (LRT) detector analytically and formulate a closed-form detection criterion. Additionally, we conduct analyses on the detector, including the computation of both the probability of false alarm and probability of detection. Simulation results demonstrate the superior efficacy of our proposed detector in comparison to existing detectors in the literature. Notably, the heightened performance is achieved without sacrificing simplicity, showcasing the added value and practicality of our approach in complex signal processing scenarios.

The remainder of the paper is as follows. Section~\ref{sec:ProblemForm} explains the system model and the problem defined in the paper. In Section~\ref{sec: prop}, the proposed detector is derived. Analysis of the proposed algorithm is outlined in Section~\ref{sec: Ana}. The simulation results are illustrated in Section~\ref{sec: Simulation} and conclusions with mentioned future works are provided in Section~\ref{sec: con}.

\section{System Model and Problem Formulation}
\label{sec:ProblemForm}
\begin{figure}[!t]
\centering
\includegraphics[width=0.25\textwidth]{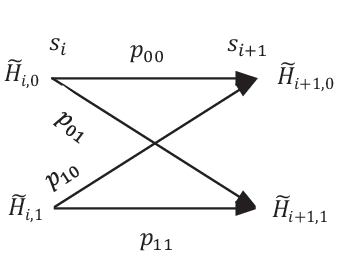}
\vspace{-0.5em}
\caption{First Order Hidden Markov Model (HMM)\label{Fig.1}}
\end{figure}

The signal vector is $\sbb=[s_1,...,s_N]^T$, the observed one bit signal is $\rb=[r_1,...,r_N]^T$, and $N$ is the length of the signal. The detection model is
\begin{equation}
\Bigg\{\begin{array}{ll}
\mathrm{H}_0:\quad\rb=\mathrm{sign}(\nb),\\
\mathrm{H}_1:\quad\rb=\mathrm{sign}(\sbb+\nb),
                  \end{array}
\end{equation}
where $\mathrm{H}_0$ is the hypothesis of the signal absence, $\mathrm{H}_1$ is the hypothesis of the signal presence, and $\nb=[n_1,...,n_N]^T$ is the noise vector. The noise vector is assumed to be White Gaussian Noise (WGN) with variance $\sigma^2$. For the sparse signal $s_i$ at time index $i$, it is assumed to be Markovian BG with two states of active state as $\tilde{\mathrm{H}}_{i,1}$ and inactive state as $\tilde{\mathrm{H}}_{i,0}$. Hence, we have $s_i\sim p_{i,0}N(0,\sigma^2_0)+(1-p_{i,0})N(0,\sigma^2_1)$. The Markovian BG model is a first order Hidden Markov Model (HMM) which is shown in Fig. \ref{Fig.1}. The transition probabilities are $p_{00}$, $p_{01}$, $p_{10}$, and $p_{11}$. It can be considered as a model for block sparse signal \cite{Habibi21}. For the correlations $\mathrm{E}\{s_is_{i+1}\}$, we assume that if both of $s_i$ and $s_{i+1}$ belongs to active states, we have $\mathrm{E}\{s_is_{i+1}\}=r\sigma_1^2$, and otherwise the successive samples are independent. The problem of sparse signal detection is to determine the hypothesis of $\mathrm{H}_0$ or $\mathrm{H}_1$ using the one bit signal vector $\rb$.

\section{The Proposed detection scheme}
\label{sec: prop}
In this section, the LRT detector is used for sparse signal detection and closed-form formula for signal detection is obtained. In this case, the presence of signal is detected if we have
\begin{equation}
\label{eq: pri9}
p(H_1)p(\rb|H_1)\ge p(H_0)p(\rb|H_0),
\end{equation}
where it is assumed that $p(H_0)=1-p(H_1)=p^{'}_0$. For the null hypothesis, $r_i$'s are independent. Therefore, it is deduced that $p(\rb|H_0)=(\frac{1}{2})^N$. To compute $p(\rb|H_1)$, since $r_i=\mathrm{sign}(s_i+n_i)$, the $r_i$'s are not independent and we have
\begin{equation}
\label{eq: pri8}
p(\rb|H_1)=p(r_1|H_1)\prod_{i=1}^{N-1}p(r_{i+1}|r_i,H_1),
\end{equation}
since there is only the first order successive dependency. It can be seen that $p(r_1|H_1)=p(s_1+n_1>0)=\frac{1}{2}$, because $s_1\sim p_{1,0}\mathcal{N}(0,\sigma^2_0)+(1-p_{1,0})\mathcal{N}(0,\sigma^2_1)$ and $n_1\sim\mathcal{N}(0,\sigma^2)$ and therefore, $s_1+n_1\sim p_{1,0}\mathcal{N}(0,\sigma^2_0+\sigma^2)+(1-p_{1,0})\mathcal{N}(0,\sigma^2_1+\sigma^2)$. To compute $p(r_{i+1}|r_i,H_1)$, we can write
\begin{IEEEeqnarray}{rCl}
p(r_{i+1}|r_i,H_1)&=&p(r_{i+1}|r_i,H_1,\tilde{H}_{i,0})p(\tilde{H}_{i,0})\nonumber\\
&+&p(r_{i+1}|r_i,H_1,\tilde{H}_{i,1})p(\tilde{H}_{i,1})\nonumber\\
&=&p(r_{i+1}|r_i,H_1,\tilde{H}_{i,0})p_{i,0}\nonumber\\
&+&p(r_{i+1}|r_i,H_1,\tilde{H}_{i,1})(1-p_{i,0}),
\label{eq: pri1}
\end{IEEEeqnarray}
where $p_{i,0}\triangleq p(\tilde{H}_{i,0})=1-p(\tilde{H}_{i,1})$. For computing $p_{i,0}$, we have
\begin{equation}
[p_{i,0}, 1-p_{i,0}]=[p_{1,0}, 1-p_{1,0}]\begin{bmatrix}
                                         1-p_{01} & p_{01} \\
                                         p_{10} & 1-p_{10}
                                       \end{bmatrix}^{i-1}.
\end{equation}
To compute $p(r_{i+1}|r_i,H_1,\tilde{H}_{i,0})$ in (\ref{eq: pri1}), we should compute it in four cases of $p(r_{i+1}=1|r_i=1,H_1,\tilde{H}_{i,0})$, $p(r_{i+1}=1|r_i=0,H_1,\tilde{H}_{i,0})$, $p(r_{i+1}=0|r_i=1,H_1,\tilde{H}_{i,0})$, and $p(r_{i+1}=0|r_i=0,H_1,\tilde{H}_{i,0})$. The first case is
\begin{IEEEeqnarray}{rCl}
p(r_{i+1}=1|r_i=1,H_1,\tilde{H}_{i,0})\;\quad\qquad\qquad\qquad\qquad\qquad\qquad\nonumber\\
=p(s_{i+1}+n_{i+1}>0|s_i+n_i>0,\tilde{H}_{i,0}).\nonumber\\
\end{IEEEeqnarray}
Because $s_i$ is inactive, $s_{i+1}$ is independent of it. Therefore, $s_{i+1}|\tilde{H}_{i,0}\sim (1-p_{01})\mathcal{N}(0,\sigma^2_0)+p_{01}\mathcal{N}(0,\sigma^2_1)$ and
\begin{IEEEeqnarray}{rCl}
p(s_{i+1}+n_{i+1}>0|s_i+n_i>0,\tilde{H}_{i,0})\;\;\quad\qquad\qquad\qquad\qquad\nonumber\\
=p(s_{i+1}+n_{i+1}>0|\tilde{H}_{i,0})=\frac{1-p_{01}}{2}+\frac{p_{01}}{2}=\frac{1}{2}.\nonumber\\
\end{IEEEeqnarray}
The other three cases are computed similarly and we have
\begin{equation}
p(r_{i+1}=1|r_i=0,H_1,\tilde{H}_{i,0})=\frac{1}{2},
\end{equation}
\begin{equation}
p(r_{i+1}=0|r_i=1,H_1,\tilde{H}_{i,0})=\frac{1}{2},
\end{equation}
\begin{equation}
\label{eq: pri12}
p(r_{i+1}=0|r_i=0,H_1,\tilde{H}_{i,0})=\frac{1}{2}.
\end{equation}
So, we can write the four cases in one equation as
\begin{equation}
\label{eq: pri5}
p(r_{i+1}|r_i,H_1,\tilde{H}_{i,0})=\frac{1}{2}.
\end{equation}
Moreover, to compute $p(r_{i+1}|r_i,H_1,\tilde{H}_{i,1})$ in (\ref{eq: pri1}), we should compute it in four cases of $p(r_{i+1}=1|r_i=1,H_1,\tilde{H}_{i,1})$, $p(r_{i+1}=1|r_i=0,H_1,\tilde{H}_{i,1})$, $p(r_{i+1}=0|r_i=1,H_1,\tilde{H}_{i,1})$, and $p(r_{i+1}=0|r_i=0,H_1,\tilde{H}_{i,1})$. The first case is
\begin{IEEEeqnarray}{rCl}
\label{eq: pri2}
p(r_{i+1}=1|r_i=1,H_1,\tilde{H}_{i,1})\;\;\quad\qquad\qquad\qquad\qquad\qquad\qquad\nonumber\\
=p(s_{i+1}+n_{i+1}>0|s_i+n_i>0,\tilde{H}_{i,1})\;\;\nonumber\\
=\frac{p(s_{i+1}+n_{i+1}>0,s_i+n_i>0|\tilde{H}_{i,1})}{p(s_i+n_i>0|\tilde{H}_{i,1})}.\nonumber\\
\end{IEEEeqnarray}
Since $s_i|\tilde{H}_{i,1}\sim\mathcal{N}(0,\sigma^2_1)$ and $n_i\sim\mathcal{N}(0,\sigma^2)$, we have $s_i+n_i|\tilde{H}_{i,1}\sim\mathcal{N}(0,\sigma^2+\sigma^2_1)$. So, we have $p(s_i+n_i>0|\tilde{H}_{i,1})=\frac{1}{2}$. In order to compute $p(s_{i+1}+n_{i+1}>0,s_i+n_i>0|\tilde{H}_{i,1})$ in (\ref{eq: pri2}), we have
\begin{IEEEeqnarray}{rCl}
p(s_{i+1}+n_{i+1}>0,s_i+n_i>0|\tilde{H}_{i,1})\quad\quad\qquad\qquad\qquad\qquad\nonumber\\
=p(s_{i+1}+n_{i+1}>0,s_i+n_i>0|\tilde{H}_{i,1},\tilde{H}_{i+1,0})\qquad\qquad\quad\nonumber\\
\times p(\tilde{H}_{i+1,0}|\tilde{H}_{i,1})\qquad\qquad\qquad\qquad\qquad\qquad\qquad\qquad\quad\nonumber\\
+p(s_{i+1}+n_{i+1}>0,s_i+n_i>0|\tilde{H}_{i,1},\tilde{H}_{i+1,1})\qquad\qquad\quad\;\nonumber\\
\times p(\tilde{H}_{i+1,1}|\tilde{H}_{i,1}),\qquad\qquad\qquad\qquad\qquad\qquad\qquad\qquad\quad
\end{IEEEeqnarray}
where $p(\tilde{H}_{i+1,0}|\tilde{H}_{i,1})=p_{10}$ and $p(\tilde{H}_{i+1,1}|\tilde{H}_{i,1})=1-p_{10}$. If $s_{i+1}$ is inactive, it is independent of $s_i$. Therefore,
\begin{IEEEeqnarray}{rCl}
p(s_{i+1}+n_{i+1}>0,s_i+n_i>0|\tilde{H}_{i,1},\tilde{H}_{i+1,0})\qquad\qquad\qquad\nonumber\\
=p(s_{i+1}+n_{i+1}>0|\tilde{H}_{i+1,0})p(s_i+n_i>0|\tilde{H}_{i,1})=\frac{1}{4}.\qquad\nonumber\\
\end{IEEEeqnarray}
To compute $p(s_{i+1}+n_{i+1}>0,s_i+n_i>0|\tilde{H}_{i,1},\tilde{H}_{i+1,1})$, the joint probability density function (PDF) $p(s_{i+1}+n_{i+1},s_i+n_i|\tilde{H}_{i,1},\tilde{H}_{i+1,1})$ is needed (since $s_i$ and $s_{i+1}$ are dependent). In what follows, we define $z_i\triangleq s_i+n_i$ and $z_{i+1}\triangleq s_{i+1}+n_{i+1}$. Because $s_i$ and $s_{i+1}$ are jointly Gaussian, $z_i$ and $z_{i+1}$ will be also jointly Gaussian. So, the jointly PDF of $z_i$ and $z_{i+1}$ is
\begin{IEEEeqnarray}{rCl}
f(z_i,z_{i+1})=\frac{1}{2\pi\sqrt{|C_2|}}\exp\{-\frac{1}{2}\zb^T_2C^{-1}_2\zb_2\},
\end{IEEEeqnarray}
where $\zb_2=[z_i, z_{i+1}]^T$, $C_2$ is the covariance matrix of $\zb_2$ and $|C_2|$ is the determinant of $C_2$. The elements of $C_2$ are $c_{2,11}=c_{2,22}=\sigma^2_1+\sigma^2$ and $c_{2,12}=c_{2,21}=r\sigma_1^2$. So, we have
\begin{IEEEeqnarray}{rCl}
f(z_i,z_{i+1})&=&\frac{1}{2\pi\sqrt{|C_2|}}\nonumber\\
&\times&\exp\big\{\frac{-1}{2|C_2|}\big((\sigma^2_1+\sigma^2)(z^2_i+z^2_{i+1})\nonumber\\
&&\qquad\qquad\quad-2r\sigma_1^2z_iz_{i+1}\big)\big\}.
\end{IEEEeqnarray}
Therefore, $p(s_{i+1}+n_{i+1}>0,s_i+n_i>0|\tilde{H}_{i,1},\tilde{H}_{i+1,1})=\int_{0}^{\infty}\int_{0}^{\infty}f(z_i,z_{i+1})\mathrm{d}z_i\mathrm{d}z_{i+1}$. If we define $p\triangleq\int_{0}^{\infty}\int_{0}^{\infty}f(z_i,z_{i+1})\mathrm{d}z_i\mathrm{d}z_{i+1}$, we will have
\begin{IEEEeqnarray}{rCl}
p(s_{i+1}+n_{i+1}>0|s_i+n_i>0,\tilde{H}_{i,1})=2p(1-p_{10})+\frac{p_{10}}{2}.\nonumber\\
\end{IEEEeqnarray}
Another case of $p(r_{i+1}|r_i,H_1,\tilde{H}_{i,1})$ that should be computed is
\begin{IEEEeqnarray}{rCl}
p(r_{i+1}=1|r_i=0,H_1,\tilde{H}_{i,1})\;\quad\quad\quad\qquad\qquad\qquad\qquad\qquad\qquad\qquad\nonumber\\
=p(s_{i+1}+n_{i+1}>0|s_i+n_i<0,\tilde{H}_{i,1})\quad\quad\qquad\qquad\qquad\qquad\qquad\nonumber\\
=\frac{p(s_{i+1}+n_{i+1}>0,s_i+n_i<0|\tilde{H}_{i,1})}{p(s_i+n_i<0|\tilde{H}_{i,1})}\quad\;\;\;\qquad\qquad\qquad\qquad\qquad\nonumber\\
=2(1-p_{10})p(s_{i+1}+n_{i+1}>0,s_i+n_i<0|\tilde{H}_{i,1},\tilde{H}_{i+1,1})\;\;\quad\quad\qquad\nonumber\\
+\frac{p_{10}}{2},\;\;\qquad\qquad\qquad\qquad\qquad\qquad\qquad\qquad\qquad\qquad\qquad\qquad\;\;
\end{IEEEeqnarray}
where $p(s_{i+1}+n_{i+1}>0,s_i+n_i<0|\tilde{H}_{i,1},\tilde{H}_{i+1,1})=\int_{0}^{\infty}\int_{-\infty}^{0}f(z_i,z_{i+1})\mathrm{d}z_i\mathrm{d}z_{i+1}$. By changing the variables in the integrals we have
\begin{IEEEeqnarray}{rCl}
\label{eq: pri3}
p(s_{i+1}+n_{i+1}>0,s_i+n_i<0|\tilde{H}_{i,1},\tilde{H}_{i+1,1})\quad\qquad\qquad\qquad\nonumber\\
=p(s_{i+1}+n_{i+1}<0,s_i+n_i>0|\tilde{H}_{i,1},\tilde{H}_{i+1,1}),\nonumber\\
\end{IEEEeqnarray}
and
\begin{IEEEeqnarray}{rCl}
\label{eq: pri4}
p(s_{i+1}+n_{i+1}>0,s_i+n_i>0|\tilde{H}_{i,1},\tilde{H}_{i+1,1})\quad\qquad\qquad\qquad\nonumber\\
=p(s_{i+1}+n_{i+1}<0,s_i+n_i<0|\tilde{H}_{i,1},\tilde{H}_{i+1,1}).\nonumber\\
\end{IEEEeqnarray}
Since the sum of the four probabilities in (\ref{eq: pri3}) and (\ref{eq: pri4}) is 1, we can write
\begin{equation}
p(s_{i+1}+n_{i+1}>0,s_i+n_i<0|\tilde{H}_{i,1},\tilde{H}_{i+1,1})=\frac{1}{2}-p.
\end{equation}
Therefore,
\begin{equation}
p(r_{i+1}=1|r_i=0,H_1,\tilde{H}_{i,1})=(1-p_{10})(1-2p)+\frac{p_{10}}{2}.
\end{equation}
Similarly, for the other two cases, it can be written
\begin{equation}
p(r_{i+1}=0|r_i=1,H_1,\tilde{H}_{i,1})=(1-p_{10})(1-2p)+\frac{p_{10}}{2},
\end{equation}
\begin{equation}
\label{eq: pri13}
p(r_{i+1}=0|r_i=0,H_1,\tilde{H}_{i,1})=2p(1-p_{10})+\frac{p_{10}}{2}.\qquad
\end{equation}
So, we can write the four cases in one equation as
\begin{equation}
\label{eq: pri6}
p(r_{i+1}|r_i,H_1,\tilde{H}_{i,1})=(1-p_{10})\hat{p}^{e_i}(1-\hat{p})^{1-e_i}+\frac{p_{10}}{2},
\end{equation}
where $\hat{p}=2p$, and
\begin{equation}
e_i=\begin{cases}
      1 & \mbox{if } r_i=r_{i+1}, \\
      0 & \mbox{if } r_i\neq r_{i+1}.
    \end{cases}
\end{equation}
Therefore, from (\ref{eq: pri1}), (\ref{eq: pri5}), and (\ref{eq: pri6}), we can write $p(r_{i+1}|r_i,H_1)$ as
\begin{IEEEeqnarray}{rCl}
p(r_{i+1}|r_i,H_1)&=&\frac{p_{i,0}}{2}+\frac{p_{10}}{2}(1-p_{i,0})\nonumber\\
&+&(1-p_{10})(1-p_{i,0})\hat{p}^{e_i}(1-\hat{p})^{1-e_i}.
\end{IEEEeqnarray}
If we define $a_i\triangleq\frac{p_{i,0}}{2}+\frac{p_{10}}{2}(1-p_{i,0})$, $p(r_{i+1}|r_i,H_1)$ can be written as
\begin{IEEEeqnarray}{rCl}
\label{eq: pri10}
p(r_{i+1}|r_i,H_1)=a_i+(1-2a_i)\hat{p}^{e_i}(1-\hat{p})^{1-e_i}.
\end{IEEEeqnarray}
From (\ref{eq: pri8}) and (\ref{eq: pri10}), the detector of the signal presence in (\ref{eq: pri9}) is
\begin{IEEEeqnarray}{rCl}
\frac{1-p_0^{'}}{2}\prod_{i=1}^{N-1}\big(a_i+(1-2a_i)\hat{p}^{e_i}(1-\hat{p})^{(1-e_i)}\big)>\frac{p_0^{'}}{2^N}.
\end{IEEEeqnarray}
Equivalently, by taking natural logarithm, we have
\begin{IEEEeqnarray}{rCl}
\label{eq: pri11}
\sum_{i=1}^{N-1}\ln\big(a_i+(1-2a_i)\hat{p}^{e_i}(1-\hat{p})^{(1-e_i)}\big)>&\ln&(\frac{p_0^{'}}{1-p_0^{'}})\nonumber\\
&-&(N-1)\ln2.\nonumber\\
\end{IEEEeqnarray}

\section{Theoretical analysis of the detector}
\label{sec: Ana}
In this section, detection probability and false alarm probability are calculated for the detector of (\ref{eq: pri11}). The detection statistic and the detection threshold are defined as $t\triangleq\sum_{i=1}^{N-1}\ln\big(a_i+(1-2a_i)\hat{p}^{e_i}(1-\hat{p})^{(1-e_i)}$ and $th\triangleq\ln(\frac{p_0^{'}}{1-p_0^{'}})-(N-1)\ln2$ respectively. Therefore, the false alarm probability ($p_{fa}$) is equal to $p_{fa}=p(t>th|H_0)$. For computing the false alarm probability, firstly, we show that $e_i$'s are independent identically distributed (IID). In fact, we have
\begin{IEEEeqnarray}{rCl}
p(e_i=1|H_0)&=&p(r_i=r_{i+1}|H_0)\nonumber\\
&=&p(r_i=r_{i+1}=0|H_0)+p(r_i=r_{i+1}=1|H_0)\nonumber\\
&=&p(\mathrm{sign}(n_i)<0,\mathrm{sign}(n_{i+1})<0)\nonumber\\
&+&p(\mathrm{sign}(n_i)>0,\mathrm{sign}(n_{i+1})>0)\nonumber\\
&=&p(\mathrm{sign}(n_i)<0)p(\mathrm{sign}(n_{i+1})<0)\nonumber\\
&+&p(\mathrm{sign}(n_i)>0)p(\mathrm{sign}(n_{i+1})>0)=\frac{1}{2}.
\end{IEEEeqnarray}
So, the distribution of $e_i$'s is not dependent on $i$ and they are identically distributed. In Appendix \ref{H0IID}, it is shown that they are also independent. Hence, $t_i=\ln\big(a_i+(1-2a_i)\hat{p}^{e_i}(1-\hat{p})^{(1-e_i)}\big)$ are also IID. Therefore, by using the central limit theorem (CLT), as $N$ goes to infinity, $t|\mathrm{H}_0$ tends to a Gaussian distribution. So, it is sufficient to find the mean ($\mu^{'}_0$) and variance ($\sigma^{'2}_0$) of the distribution as
\begin{IEEEeqnarray}{rCl}
\mu^{'}_0&=&\mathrm{E}\{t|H_0\}\nonumber\\
&=&\frac{1}{2}\sum_{i=1}^{N-1}\Big\{\ln\big(a_i+(1-2a_i)\hat{p}\big)\nonumber\\
&+&\ln\big(a_i+(1-2a_i)(1-\hat{p})\big)\Big\}.
\end{IEEEeqnarray}
If we define $c_i\triangleq a_i+(1-2a_i)\hat{p}$, we have
\begin{IEEEeqnarray}{rCl}
\mu^{'}_0=\frac{1}{2}\sum_{i=1}^{N-1}\big(\ln c_i+\ln(1-c_i)\big).
\end{IEEEeqnarray}
In order to compute $\sigma^{'2}_0$, at first, we should compute $\mathrm{E}\{t^2|H_0\}$ as
\begin{IEEEeqnarray}{rCl}
\mathrm{E}\{&t^2&|H_0)\}=\sum_{i=1}^{N-1}\mathrm{E}\Big\{\ln^2\big(a_i+(1-2a_i)\hat{p}^{e_i}(1-\hat{p})^{1-e_i}\big)\Big\}\nonumber\\
&+&\underset{i\neq j}{\sum_{i=1}^{N-1}\sum_{j=1}^{N-1}}\,\bigg(\mathrm{E}\Big\{\ln\big(a_i+(1-2a_i)\hat{p}^{e_i}(1-\hat{p})^{1-e_i}\big)\Big\}\nonumber\\
&\times&\mathrm{E}\Big\{\ln\big(a_j+(1-2a_j)\hat{p}^{e_j}(1-\hat{p})^{1-e_j}\big)\Big\}\bigg)\nonumber\\
&=&\frac{1}{2}\sum_{i=1}^{N-1}\big(\ln^2c_i+\ln^2(1-c_i)\big)\nonumber\\
&+&\frac{1}{4}\underset{i\neq j}{\sum_{i=1}^{N-1}\sum_{j=1}^{N-1}}\,\big(\ln c_i+\ln(1-c_i)\big)\big(\ln c_j+\ln(1-c_j)\big).\nonumber\\
\end{IEEEeqnarray}
Therefore, $\sigma^{'2}_0$ is
\begin{IEEEeqnarray}{rCl}
\sigma^{'2}_0=\mathrm{E}\{t^2|H_0\}-\mu^{'2}_0=\frac{1}{4}\sum_{i=1}^{N-1}\big(\ln c_i-\ln(1-c_i)\big)^2.\nonumber\\
\end{IEEEeqnarray}
So, the false alarm probability can be written as
\begin{IEEEeqnarray}{rCl}
p_{fa}=\mathrm{Q}\Big(\frac{th-\mu^{'}_0}{\sigma^{'}_0}\Big),
\end{IEEEeqnarray}
where $\mathrm{Q}(.)$ is the Gaussian Q function.\\
\indent Now, we want to calculate the detection probability. The detection probability ($p_d$) is equal to $p_d=p(t>th|\mathrm{H_1})$. In order to find $p_d$, at first, we should compute $p(e_i|\mathrm{H_1})$ as
\begin{IEEEeqnarray}{rCl}
p(e_i=1|H_1)&=&p(r_i=r_{i+1}|H_1)\nonumber\\
&=&p(r_i=r_{i+1}=0|H_1)+p(r_i=r_{i+1}=1|H_1).\nonumber\\
\end{IEEEeqnarray}
To compute $p(r_i=r_{i+1}=0|H_1)$, we have
\begin{IEEEeqnarray}{rCl}
p(&r_i&=r_{i+1}=0|H_1)=p(s_i+n_i<0,s_{i+1}+n_{i+1}<0)\qquad\nonumber\\
&=&p(s_i+n_i<0,s_{i+1}+n_{i+1}<0|\tilde{H}_{i,0})p_{i,0}\nonumber\\
&+&p(s_i+n_i<0,s_{i+1}+n_{i+1}<0|\tilde{H}_{i,1})(1-p_{i,0})\nonumber\\
&=&p(s_{i+1}+n_{i+1}<0|s_i+n_i<0,\tilde{H}_{i,0})\nonumber\\
&\times&p(s_i+n_i<0|\tilde{H}_{i,0})p_{i,0}\nonumber\\
&+&p(s_{i+1}+n_{i+1}<0|s_i+n_i<0,\tilde{H}_{i,1})\nonumber\\
&\times&p(s_i+n_i<0|\tilde{H}_{i,1})(1-p_{i,0})\nonumber\\
&=&\frac{p_{i,0}}{2}p(r_{i+1}=0|r_i=0,H_1,\tilde{H}_{i,0})\nonumber\\
&+&\frac{1-p_{i,0}}{2}p(r_{i+1}=0|r_i=0,H_1,\tilde{H}_{i,1})\nonumber\\
&=&\frac{p_{i,0}}{4}+\frac{1-p_{i,0}}{2}\big(2p(1-p_{10})+\frac{p_{10}}{2}\big),
\end{IEEEeqnarray}
where we have used (\ref{eq: pri12}) and (\ref{eq: pri13}). Similarly, to compute $p(r_i=r_{i+1}=1|H_1)$, we have
\begin{IEEEeqnarray}{rCl}
p(&r_i&=r_{i+1}=1|H_1)=p(s_i+n_i>0,s_{i+1}+n_{i+1}>0)\qquad\nonumber\\
&=&p(s_i+n_i>0,s_{i+1}+n_{i+1}>0|\tilde{H}_{i,0})p_{i,0}\nonumber\\
&+&p(s_i+n_i>0,s_{i+1}+n_{i+1}>0|\tilde{H}_{i,1})(1-p_{i,0})\nonumber\\
&=&p(s_{i+1}+n_{i+1}>0|s_i+n_i>0,\tilde{H}_{i,0})\nonumber\\
&\times&p(s_i+n_i>0|\tilde{H}_{i,0})p_{i,0}\nonumber\\
&+&p(s_{i+1}+n_{i+1}>0|s_i+n_i>0,\tilde{H}_{i,1})\nonumber\\
&\times&p(s_i+n_i>0|\tilde{H}_{i,1})(1-p_{i,0})\nonumber\\
&=&\frac{p_{i,0}}{2}p(r_{i+1}=1|r_i=1,H_1,\tilde{H}_{i,0})\nonumber\\
&+&\frac{1-p_{i,0}}{2}p(r_{i+1}=1|r_i=1,H_1,\tilde{H}_{i,1})\nonumber\\
&=&\frac{p_{i,0}}{4}+\frac{1-p_{i,0}}{2}\big(2p(1-p_{10})+\frac{p_{10}}{2}\big).
\end{IEEEeqnarray}
Therefore, $p(e_i=1|\mathrm{H_1})$ equals to
\begin{IEEEeqnarray}{rCl}
p(e_i=1|H_1)&=&\frac{p_{i,0}}{2}+(1-p_{i,0})\big(2p(1-p_{10})+\frac{p_{10}}{2}\big)\nonumber\\
&=&c_i.
\end{IEEEeqnarray}
As it is observed, the probability densities of $e_i$'s conditioned on $\mathrm{H}_1$ are not identical. In fact, they depend on $i$. Therefore, $e_i$'s conditioned on $\mathrm{H}_1$ are not IID. If we assume that as $N$ goes to infinity, $t|\mathrm{H}_1$ tends to a Gaussian distribution, it is sufficient to find the mean ($\mu^{'}_1$) and variance ($\sigma^{'2}_1$) of the distribution as
\begin{IEEEeqnarray}{rCl}
\mu^{'}_1=\mathrm{E}\{t|H_1\}=\sum_{i=1}^{N-1}\big(c_i\ln c_i+(1-c_i)\ln(1-c_i)\big),\\
\sigma^{'2}_1=\mathrm{E}\{t^2|H_1\}-\mu^{'2}_1,\qquad\qquad\qquad
\end{IEEEeqnarray}
where
\begin{IEEEeqnarray}{rCl}
\mathrm{E}\{t^2|H_1\}&=&\sum_{i=1}^{N-1}\mathrm{E}\Big\{\ln^2\big(a_i+(1-2a_i)\hat{p}^{e_i}(1-\hat{p})^{1-e_i}\big)\Big\}\nonumber\\
&+&\underset{i\neq j}{\sum_{i=1}^{N-1}\sum_{j=1}^{N-1}}\,\mathrm{E}\Big\{\ln\big(a_i+(1-2a_i)\hat{p}^{e_i}(1-\hat{p})^{1-e_i}\big)\nonumber\\
&\times&\ln\big(a_j+(1-2a_j)\hat{p}^{e_j}(1-\hat{p})^{1-e_j}\big)\Big\}\nonumber\\
&=&\sum_{i=1}^{N-1}\big(c_i\ln^2c_i+(1-c_i)\ln^2(1-c_i)\big)\nonumber\\
&+&\underset{i\neq j}{\sum_{i=1}^{N-1}\sum_{j=1}^{N-1}}\,\big(p^{'}_{00ij}\ln(1-c_i)\ln(1-c_j)\nonumber\\
&+&p^{'}_{11ij}\ln c_i\ln c_j+p^{'}_{01ij}\ln(1-c_i)\ln c_j\nonumber\\
&+&p^{'}_{10ij}\ln c_i\ln(1-c_j)\big),
\end{IEEEeqnarray}
where
\begin{IEEEeqnarray}{rCl}
p^{'}_{00ij}=p(e_i=0,e_j=0|H_1),\\
p^{'}_{11ij}=p(e_i=1,e_j=1|H_1),\\
p^{'}_{01ij}=p(e_i=0,e_j=1|H_1),\\
p^{'}_{10ij}=p(e_i=1,e_j=0|H_1).
\end{IEEEeqnarray}
The probability $p^{'}_{00ij}$ is computed in Appendix \ref{CPP}. The other three probabilities can be computed in a similar manner. Finally, the detection probability can be written as
\begin{IEEEeqnarray}{rCl}
p_d=\mathrm{Q}\Big(\frac{th-\mu^{'}_1}{\sigma^{'}_1}\Big).
\end{IEEEeqnarray}


\begin{figure}[!t]
\centering
\includegraphics[width=0.4\textwidth]{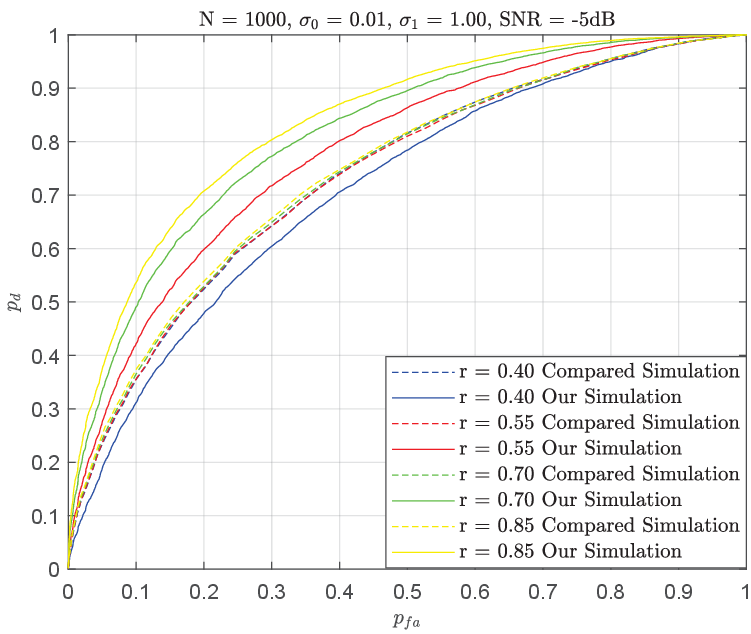}
\vspace{-0.5em}
\caption{The ROC curves for different values of correlation coefficient ($r$) at $\mathrm{SNR=-5dB}$\label{Fig.2}}
\end{figure}

\section{Simulation Results}
\label{sec: Simulation}
In this section, four simulations are conducted to experimentally evaluate the performance of the proposed detector and also compare it with the detector given in \cite{Wang19_1} with one bit quantization. In all of the simulations, the following parameters are fixed
\begin{IEEEeqnarray}{rCl}
N=1000&,&p_{1,0}=0.95,p_{10}=0.1,p_{01}=\frac{0.01}{0.9},\sigma_0=0.01\nonumber\\
&,&\sigma_1=1.
\end{IEEEeqnarray}
The signal to noise ratio (SNR) is defined by
\begin{IEEEeqnarray}{rCl}
\mathrm{SNR}=\frac{\mathrm{E}\{\sbb^T\sbb\}}{\mathrm{E}\{\nb^T\nb\}},
\end{IEEEeqnarray}
where
\begin{IEEEeqnarray}{rCl}
\mathrm{E}\{\nb^T\nb\}=N\sigma^2,
\end{IEEEeqnarray}
and
\begin{IEEEeqnarray}{rCl}
\mathrm{E}\{\sbb^T\sbb\}&=&\mathrm{E}\Big\{\sum_{i=1}^{N}s_i^2\Big\}=\sum_{i=1}^{N}\mathrm{E}\{s_i^2\}\qquad\qquad\qquad\qquad\qquad\nonumber\\
&=&\sum_{i=1}^{N}\big(\mathrm{E}\{s_i^2|\tilde{H}_{i,0}\}p(\tilde{H}_{i,0})+\mathrm{E}\{s_i^2|\tilde{H}_{i,1}\}p(\tilde{H}_{i,1})\big)\nonumber\\
&=&\sum_{i=1}^{N}\big((\sigma_0^2p(\tilde{H}_{i,0})+\sigma_1^2p(\tilde{H}_{i,1})\big)\nonumber\\
&=&\sigma_0^2\sum_{i=1}^{N}p(\tilde{H}_{i,0})+\sigma_1^2\sum_{i=1}^{N}p(\tilde{H}_{i,1})\nonumber\\
&=&\sigma_0^2\sum_{i=1}^{N}p(\tilde{H}_{i,0})+\sigma_1^2\sum_{i=1}^{N}(1-p(\tilde{H}_{i,0}))\nonumber\\
&=&\sigma_0^2\sum_{i=1}^{N}p(\tilde{H}_{i,0})+\sigma_1^2\big(N-\sum_{i=1}^{N}p(\tilde{H}_{i,0})\big)\nonumber\\
&=&N\sigma_1^2+(\sigma_0^2-\sigma_1^2)\sum_{i=1}^{N}p(\tilde{H}_{i,0}).
\end{IEEEeqnarray}
Therefore, SNR equals to
\begin{IEEEeqnarray}{rCl}
\mathrm{SNR}=\frac{\sigma_1^2+\frac{\sigma_0^2-\sigma_1^2}{N}\sum_{i=1}^{N}p(\tilde{H}_{i,0})}{\sigma^2}.
\end{IEEEeqnarray}
In the first experiment, the receiver operating characteristic (ROC) curves are plotted empirically for different values of correlation coefficient ($r$) at $\mathrm{SNR=-5dB}$ in Fig. \ref{Fig.2}. As is expected, the performance of the compared detector does not vary greatly with correlation coefficient. On the other hand, for the proposed detector, as the correlation coefficient increases, the performance of the detector improves greatly. As is observed, for $r\geq0.55$, the proposed detector performs much better than the compared detector.\\

\begin{figure}[!t]
\centering
\includegraphics[width=0.4\textwidth]{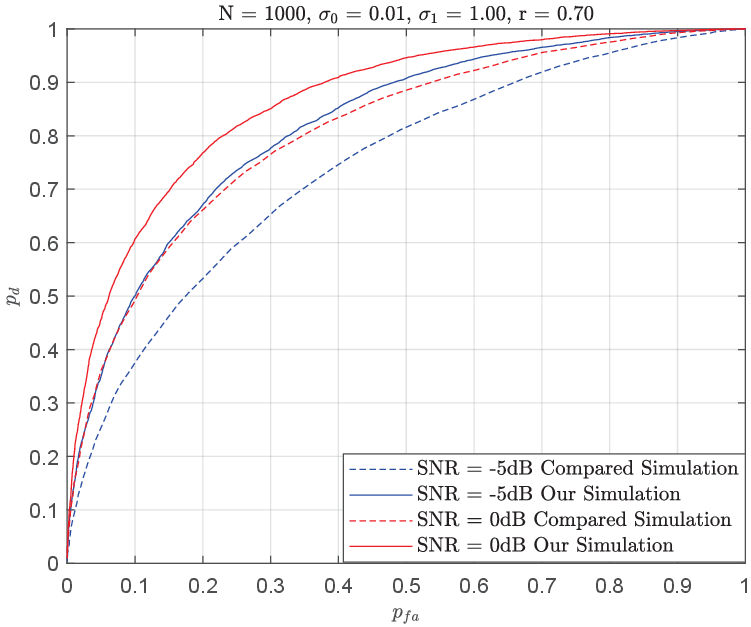}
\vspace{-0.5em}
\caption{The ROC curves for different values of SNR at $r=0.7$\label{Fig.3}}
\end{figure}

\begin{figure}[!t]
\centering
\includegraphics[width=0.4\textwidth]{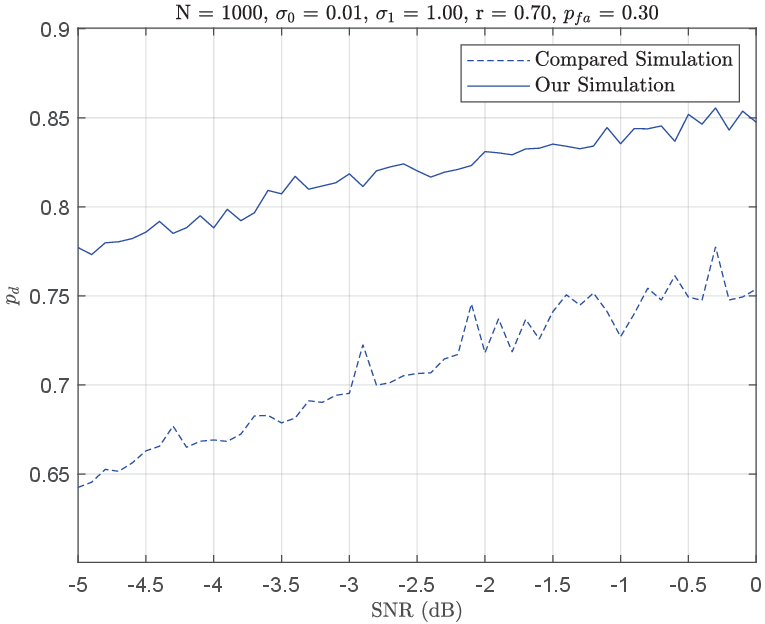}
\vspace{-0.5em}
\caption{The power function at $r=0.7$ and $p_{fa}=0.3$\label{Fig.4}}
\end{figure}

\begin{figure}[!t]
\centering
\includegraphics[width=0.4\textwidth]{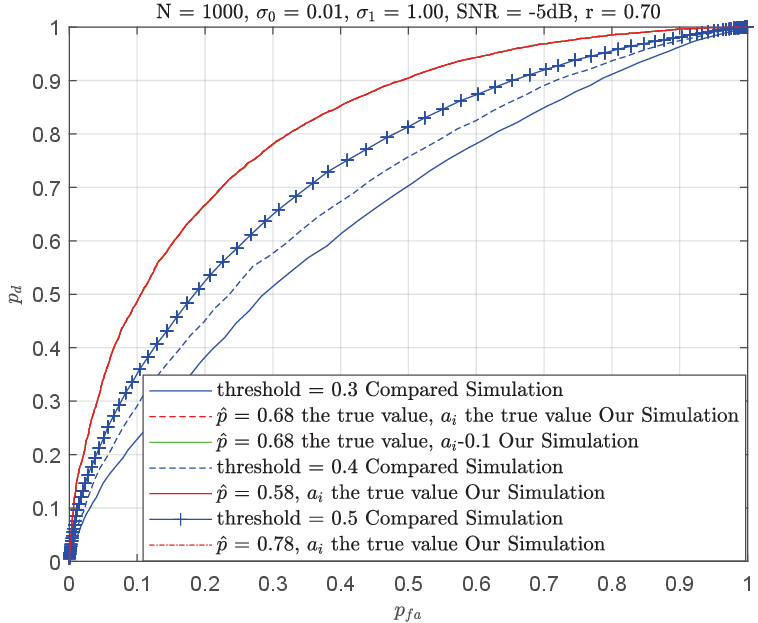}
\vspace{-0.5em}
\caption{The detectors sensitivity to their parameters at $\mathrm{SNR=-5dB}$ and $r=0.7$\label{Fig.5}}
\end{figure}

In the second experiment, the ROC curves are plotted empirically for different values of SNR at $r=0.7$ in Fig. \ref{Fig.3}. As is expected, by increasing the SNR, the performance of both detectors improves. However, as can be seen, even at the low SNR $-5$dB, the proposed detector performs at least a little better than the compared detector at SNR $0$dB.\\
In the third experiment, the power functions of both detectors ($p_d$ versus SNR) are plotted empirically at $r=0.7$ and $p_{fa}=0.3$ in Fig. \ref{Fig.4}. As is expected, by increasing the SNR, the performance of both detectors improves overall. However, as is observed, the proposed detector greatly outperforms the compared detector.\\
In the last experiment, the sensitivity of both detectors to their parameters is investigated. For the proposed detector, the parameters are $\hat{p}$ and $a_i$'s. However, for the compared detector, there is only one threshold. In Fig. \ref{Fig.5}, the ROC curves of both detectors are plotted at $\mathrm{SNR=-5dB}$ and $r=0.7$ for different values of their parameters. From Fig. \ref{Fig.5}, it is observed that, by changing the threshold of the compared detector, its performance varies greatly. However, by changing different parameters of the proposed detector, its performance variation is negligible.

%




\section{Conclusion and future work}
\label{sec: con}
In this paper, a Bayesian detector of block sparse signal using one bit measurements is derived. The block sparse signal is modeled by a correlated Hidden Markov model. The closed-form detection criterion is obtained mathematically which is shown in the simulation results that are not very sensitive to model parameters. The mathematical analysis of the detector are provided which consist of calculating the detection probability and false alarm probability. In addition to low sensitivity of the detector to model parameters, simulation results show the effectiveness of the proposed detector in comparison to other detector in the literature with some added complexity of the detector. 

\appendices
\section{Independence of $e_i$'s for the null hypothesis}
\label{H0IID}
In order to show that $e_i$'s are independent for the null hypothesis, we prove that $p(e_i,e_j|H_0)=p(e_i|H_0)p(e_j|H_0),i\neq j$. To compute $p(e_i,e_j|H_0)$, we should compute it in four cases of $p(e_i=1,e_j=1|H_0)$, $p(e_i=1,e_j=0|H_0)$, $p(e_i=0,e_j=1|H_0)$, and $p(e_i=0,e_j=0|H_0)$. The first case is
\begin{IEEEeqnarray}{rCl}
p(e_i=1,e_j=1|H_0)=p(r_i=r_{i+1},r_j=r_{j+1}|H_0).
\end{IEEEeqnarray}
We will consider three cases. Firstly, if $j=i-1$, we will have
\begin{IEEEeqnarray}{rCl}
p(&e_i&=1,e_{i-1}=1|H_0)=p(r_{i-1}=r_i=r_{i+1}|H_0)\qquad\qquad\nonumber\\
&=&p(r_{i-1}=r_i=r_{i+1}=0|H_0)\nonumber\\
&+&p(r_{i-1}=r_i=r_{i+1}=1|H_0)\nonumber\\
&=&p(n_i<0,n_{i-1}<0,n_{i+1}<0)\nonumber\\
&+&p(n_i>0,n_{i-1}>0,n_{i+1}>0)\nonumber\\
&=&p(n_i<0)p(n_{i-1}<0)p(n_{i+1}<0)\nonumber\\
&+&p(n_i>0)p(n_{i-1}>0)p(n_{i+1}>0)=\frac{1}{4}\nonumber\\
&=&p(e_i=1|H_0)p(e_{i-1}=1|H_0).
\end{IEEEeqnarray}
Secondly, if $j=i+1$, by symmetry we can write
\begin{IEEEeqnarray}{rCl}
p(e_i=1,e_{i+1}=1|H_0)&=&p(r_i=r_{i+1}=r_{i+2}|H_0)=\frac{1}{4}\nonumber\\
&=&p(e_i=1|H_0)p(e_{i+1}=1|H_0).\nonumber\\
\end{IEEEeqnarray}
Thirdly, if $j\neq i-1,i+1$, we have
\begin{IEEEeqnarray}{rCl}
p(&e_i&=1,e_j=1|H_0)=p(r_i=r_{i+1},r_j=r_{j+1}|H_0)\qquad\qquad\nonumber\\
&=&p(r_i=r_{i+1}=1,r_j=r_{j+1}=1|H_0)\nonumber\\
&+&p(r_i=r_{i+1}=0,r_j=r_{j+1}=0|H_0)\nonumber\\
&+&p(r_i=r_{i+1}=0,r_j=r_{j+1}=1|H_0)\nonumber\\
&+&p(r_i=r_{i+1}=1,r_j=r_{j+1}=0|H_0).
\end{IEEEeqnarray}
For $p(r_i=r_{i+1}=1,r_j=r_{j+1}=1|H_0)$, we can write
\begin{IEEEeqnarray}{rCl}
p(&r_i&=r_{i+1}=1,r_j=r_{j+1}=1|H_0)\qquad\qquad\qquad\qquad\quad\nonumber\\
&=&p(n_i>0,n_{i+1}>0,n_j>0,n_{j+1}>0)\nonumber\\
&=&p(n_i>0)p(n_{i+1}>0)p(n_j>0)p(n_{j+1}>0)=\frac{1}{16}.\nonumber\\
\end{IEEEeqnarray}
Similarly, we can write
\begin{IEEEeqnarray}{rCl}
p(&r_i&=r_{i+1}=0,r_j=r_{j+1}=0|H_0)\qquad\qquad\qquad\qquad\quad\nonumber\\
&=&p(r_i=r_{i+1}=1,r_j=r_{j+1}=0|H_0)\nonumber\\
&=&p(r_i=r_{i+1}=0,r_j=r_{j+1}=1|H_0)=\frac{1}{16}.
\end{IEEEeqnarray}
Therefore, $p(e_i=1,e_j=1|H_0)$ will be
\begin{IEEEeqnarray}{rCl}
p(e_i=1,e_j=1|H_0)&=&p(r_i=r_{i+1},r_j=r_{j+1}|H_0)=\frac{1}{4}\nonumber\\
&=&p(r_i=r_{i+1}|H_0)p(r_j=r_{j+1}|H_0)\nonumber\\
&=&p(e_i=1|H_0)p(e_j=1|H_0).
\end{IEEEeqnarray}
The second case is
\begin{IEEEeqnarray}{rCl}
p(e_i=0,e_j=0|H_0)=p(r_i\neq r_{i+1},r_j\neq r_{j+1}|H_0).
\end{IEEEeqnarray}
Again, we will consider three cases. Firstly, if $j=i-1$, we will have
\begin{IEEEeqnarray}{rCl}
p(&e_i&=0,e_{i-1}=0|H_0)=p(r_{i-1}\neq r_i, r_i\neq r_{i+1}|H_0)\qquad\nonumber\\
&=&p(r_i=0,r_{i-1}=r_{i+1}=1|H_0)\nonumber\\
&+&p(r_i=1,r_{i-1}=r_{i+1}=0|H_0)\nonumber\\
&=&p(n_i<0,n_{i-1}>0,n_{i+1}>0)\nonumber\\
&+&p(n_i>0,n_{i-1}<0,n_{i+1}<0)=\frac{1}{4}\nonumber\\
&=&p(e_i=0|H_0)p(e_{i-1}=0|H_0).
\end{IEEEeqnarray}
Secondly, if $j=i+1$, by symmetry we can write
\begin{IEEEeqnarray}{rCl}
p(e_i=0,e_{i+1}=0|H_0)&=&p(r_i\neq r_{i+1},r_{i+1}\neq r_{i+2}|H_0)=\frac{1}{4}\nonumber\\
&=&p(e_i=0|H_0)p(e_{i+1}=0|H_0).
\end{IEEEeqnarray}
Thirdly, if $j\neq i-1,i+1$, we have
\begin{IEEEeqnarray}{rCl}
p(&e_i&=0,e_j=0|H_0)=p(r_i\neq r_{i+1},r_j\neq r_{j+1}|H_0)\quad\qquad\nonumber\\
&=&p(r_i=0,r_{i+1}=1,r_j=0,r_{j+1}=1|H_0)\nonumber\\
&+&p(r_i=0,r_{i+1}=1,r_j=1,r_{j+1}=0|H_0)\nonumber\\
&+&p(r_i=1,r_{i+1}=0,r_j=0,r_{j+1}=1|H_0)\nonumber\\
&+&p(r_i=1,r_{i+1}=0,r_j=1,r_{j+1}=0|H_0).
\end{IEEEeqnarray}
For $p(r_i=0,r_{i+1}=1,r_j=0,r_{j+1}=1|H_0)$, we can write
\begin{IEEEeqnarray}{rCl}
p(&r_i&=0,r_{i+1}=1,r_j=0,r_{j+1}=1|H_0)\qquad\qquad\qquad\quad\nonumber\\
&=&p(n_i<0,n_{i+1}>0,n_j<0,n_{j+1}>0)=\frac{1}{16}.
\end{IEEEeqnarray}
Similarly, we can write
\begin{IEEEeqnarray}{rCl}
p(&r_i&=0,r_{i+1}=1,r_j=1,r_{j+1}=0|H_0)\qquad\qquad\qquad\quad\nonumber\\
&=&p(r_i=1,r_{i+1}=0,r_j=0,r_{j+1}=1|H_0)\nonumber\\
&=&p(r_i=1,r_{i+1}=0,r_j=1,r_{j+1}=0|H_0)=\frac{1}{16}.
\end{IEEEeqnarray}
Therefore, $p(e_i=0,e_j=0|H_0)$ will be
\begin{IEEEeqnarray}{rCl}
p(e_i=0,e_j=0|H_0)&=&p(r_i\neq r_{i+1},r_j\neq r_{j+1}|H_0)=\frac{1}{4}\nonumber\\
&=&p(r_i\neq r_{i+1}|H_0)p(r_j\neq r_{j+1}|H_0)\nonumber\\
&=&p(e_i=0|H_0)p(e_j=0|H_0).
\end{IEEEeqnarray}
The third case is
\begin{IEEEeqnarray}{rCl}
p(e_i=0,e_j=1|H_0)=p(r_i\neq r_{i+1},r_j=r_{j+1}|H_0).
\end{IEEEeqnarray}
Again, we will consider three cases. Firstly, if $j=i-1$, we will have
\begin{IEEEeqnarray}{rCl}
p(&e_i&=0,e_{i-1}=1|H_0)=p(r_{i-1}=r_i, r_i\neq r_{i+1}|H_0)\qquad\nonumber\\
&=&p(r_{i-1}=0,r_i=0,r_{i+1}=1|H_0)\nonumber\\
&+&p(r_{i-1}=1,r_i=1,r_{i+1}=0|H_0)\nonumber\\
&=&p(n_{i-1}<0,n_{i}<0,n_{i+1}>0)\nonumber\\
&+&p(n_{i-1}>0,n_{i}>0,n_{i+1}<0)=\frac{1}{4}\nonumber\\
&=&p(e_i=0|H_0)p(e_{i-1}=1|H_0).
\end{IEEEeqnarray}
Secondly, if $j=i+1$, we can write
\begin{IEEEeqnarray}{rCl}
p(&e_i&=0,e_{i+1}=1|H_0)=p(r_i\neq r_{i+1}, r_{i+1}=r_{i+2}|H_0)\nonumber\\
&=&p(r_i=0,r_{i+1}=1,r_{i+2}=1|H_0)\nonumber\\
&+&p(r_i=1,r_{i+1}=0,r_{i+2}=0|H_0)\nonumber\\
&=&p(n_{i}<0,n_{i+1}>0,n_{i+2}>0)\nonumber\\
&+&p(n_{i}>0,n_{i+1}<0,n_{i+2}<0)=\frac{1}{4}\nonumber\\
&=&p(e_i=0|H_0)p(e_{i+1}=1|H_0).
\end{IEEEeqnarray}
Thirdly, if $j\neq i-1,i+1$, we have
\begin{IEEEeqnarray}{rCl}
p(&e_i&=0,e_j=1|H_0)=p(r_i\neq r_{i+1},r_j=r_{j+1}|H_0)\quad\qquad\nonumber\\
&=&p(r_i=0,r_{i+1}=1,r_j=0,r_{j+1}=0|H_0)\nonumber\\
&+&p(r_i=0,r_{i+1}=1,r_j=1,r_{j+1}=1|H_0)\nonumber\\
&+&p(r_i=1,r_{i+1}=0,r_j=0,r_{j+1}=0|H_0)\nonumber\\
&+&p(r_i=1,r_{i+1}=0,r_j=1,r_{j+1}=1|H_0).
\end{IEEEeqnarray}
For $p(r_i=0,r_{i+1}=1,r_j=0,r_{j+1}=0|H_0)$, we can write
\begin{IEEEeqnarray}{rCl}
p(&r_i&=0,r_{i+1}=1,r_j=0,r_{j+1}=0|H_0)\qquad\qquad\qquad\quad\nonumber\\
&=&p(n_i<0,n_{i+1}>0,n_j<0,n_{j+1}<0)=\frac{1}{16}.
\end{IEEEeqnarray}
Similarly, we can write
\begin{IEEEeqnarray}{rCl}
p(&r_i&=0,r_{i+1}=1,r_j=1,r_{j+1}=1|H_0)\qquad\qquad\qquad\quad\nonumber\\
&=&p(r_i=1,r_{i+1}=0,r_j=0,r_{j+1}=0|H_0)\nonumber\\
&=&p(r_i=1,r_{i+1}=0,r_j=1,r_{j+1}=1|H_0)=\frac{1}{16}.
\end{IEEEeqnarray}
Therefore, $p(e_i=0,e_j=1|H_0)$ will be
\begin{IEEEeqnarray}{rCl}
p(e_i=0,e_j=1|H_0)&=&p(r_i\neq r_{i+1},r_j=r_{j+1}|H_0)=\frac{1}{4}\nonumber\\
&=&p(r_i\neq r_{i+1}|H_0)p(r_j=r_{j+1}|H_0)\nonumber\\
&=&p(e_i=0|H_0)p(e_j=1|H_0).
\end{IEEEeqnarray}
The fourth case is
\begin{IEEEeqnarray}{rCl}
p(e_i=1,e_j=0|H_0)=p(r_i=r_{i+1},r_j\neq r_{j+1}|H_0).\nonumber\\
\end{IEEEeqnarray}
By symmetry to the third case, we will have
\begin{IEEEeqnarray}{rCl}
p(e_i=1,e_j=0|H_0)&=&p(r_i=r_{i+1},r_j\neq r_{j+1}|H_0)=\frac{1}{4}\nonumber\\
&=&p(r_i=r_{i+1}|H_0)p(r_j\neq r_{j+1}|H_0)\nonumber\\
&=&p(e_i=1|H_0)p(e_j=0|H_0).
\end{IEEEeqnarray}
\section{Computation of the probability $p^{'}_{00ij}$}
\label{CPP}
In order to compute $p^{'}_{00ij}$, we have
\begin{IEEEeqnarray}{rCl}
p^{'}_{00ij}=p(e_i=0,e_j=0|H_1)=p(r_i\neq r_{i+1},r_j\neq r_{j+1}|H_1).\nonumber\\
\end{IEEEeqnarray}
Because of symmetry, without loss of generality, we can assume that $j>i$. We will consider three cases: $j=i+1$, $j=i+2$, and $j\geq i+3$. For the first case $j=i+1$, we have
\begin{IEEEeqnarray}{rCl}
p^{'}_{00i(i+1)}&=&p(r_i\neq r_{i+1},r_{i+1}\neq r_{i+2}|H_1)\nonumber\\
&=&p(r_i=0,r_{i+1}=1,r_{i+2}=0|H_1)\nonumber\\
&+&p(r_i=1,r_{i+1}=0,r_{i+2}=1|H_1).
\end{IEEEeqnarray}
We will compute $p(r_i=0,r_{i+1}=1,r_{i+2}=0|H_1)$. $p(r_i=1,r_{i+1}=0,r_{i+2}=1|H_1)$ can also be computed similarly. For $p(r_i=0,r_{i+1}=1,r_{i+2}=0|H_1)$, we have
\begin{IEEEeqnarray}{rCl}
p(&r_i&=0,r_{i+1}=1,r_{i+2}=0|H_1)\quad\qquad\qquad\qquad\qquad\qquad\nonumber\\
&=&p(s_i+n_i<0,s_{i+1}+n_{i+1}>0,s_{i+2}+n_{i+2}<0).
\end{IEEEeqnarray}
If we define $A\triangleq s_i+n_i<0,s_{i+1}+n_{i+1}>0,s_{i+2}+n_{i+2}<0$, $p(r_i=0,r_{i+1}=1,r_{i+2}=0|H_1)$ can be written as
\begin{IEEEeqnarray}{rCl}
p(A)=p(\tilde{H}_{i,0})p(A|\tilde{H}_{i,0})+(1-p(\tilde{H}_{i,0}))p(A|\tilde{H}_{i,1}).
\end{IEEEeqnarray}
For $p(A|\tilde{H}_{i,0})$, we have
\begin{IEEEeqnarray}{rCl}
p(A|\tilde{H}_{i,0})&=&p(s_i+n_i<0|\tilde{H}_{i,0})\nonumber\\
&\times&p(s_{i+1}+n_{i+1}>0,s_{i+2}+n_{i+2}<0|\tilde{H}_{i,0})\nonumber\\
&=&\frac{1}{2}p(B|\tilde{H}_{i,0}),
\end{IEEEeqnarray}
where $B\triangleq s_{i+1}+n_{i+1}>0,s_{i+2}+n_{i+2}<0$. For $p(B|\tilde{H}_{i,0})$, we have
\begin{IEEEeqnarray}{rCl}
p(B|\tilde{H}_{i,0})&=&p(\tilde{H}_{i+1,0}|\tilde{H}_{i,0})p(B|\tilde{H}_{i,0},\tilde{H}_{i+1,0})\nonumber\\
&+&p(\tilde{H}_{i+1,1}|\tilde{H}_{i,0})p(B|\tilde{H}_{i,0},\tilde{H}_{i+1,1})\nonumber\\
&=&\frac{1-p_{01}}{4}+p_{01}p(B|\tilde{H}_{i,0},\tilde{H}_{i+1,1}).
\end{IEEEeqnarray}
For $p(B|\tilde{H}_{i,0},\tilde{H}_{i+1,1})$, we have
\begin{IEEEeqnarray}{rCl}
p(B|\tilde{H}_{i,0},\tilde{H}_{i+1,1})&=&p(\tilde{H}_{i+2,0}|\tilde{H}_{i+1,1})p(B|\tilde{H}_{i+1,1},\tilde{H}_{i+2,0})\nonumber\\
&+&p(\tilde{H}_{i+2,1}|\tilde{H}_{i+1,1})p(B|\tilde{H}_{i+1,1},\tilde{H}_{i+2,1})\nonumber\\
&=&\frac{p_{10}}{4}+(1-p_{10})(\frac{1}{2}-p).
\end{IEEEeqnarray}
For $p(A|\tilde{H}_{i,1})$, we have
\begin{IEEEeqnarray}{rCl}
p(A|\tilde{H}_{i,1})&=&p(\tilde{H}_{i+1,0}|\tilde{H}_{i,1})p(A|\tilde{H}_{i,1},\tilde{H}_{i+1,0})\nonumber\\
&+&p(\tilde{H}_{i+1,1}|\tilde{H}_{i,1})p(A|\tilde{H}_{i,1},\tilde{H}_{i+1,1})\nonumber\\
&=&\frac{p_{10}}{8}+(1-p_{10})p(A|\tilde{H}_{i,1},\tilde{H}_{i+1,1}).
\end{IEEEeqnarray}
For $p(A|\tilde{H}_{i,1},\tilde{H}_{i+1,1})$, we have
\begin{IEEEeqnarray}{rCl}
p(&A&|\tilde{H}_{i,1},\tilde{H}_{i+1,1})\nonumber\\
&=&p(\tilde{H}_{i+2,0}|\tilde{H}_{i+1,1})p(A|\tilde{H}_{i,1},\tilde{H}_{i+1,1},\tilde{H}_{i+2,0})\nonumber\\
&+&p(\tilde{H}_{i+2,1}|\tilde{H}_{i+1,1})p(A|\tilde{H}_{i,1},\tilde{H}_{i+1,1},\tilde{H}_{i+2,1})\nonumber\\
&=&\frac{p_{10}}{2}(\frac{1}{2}-p)+(1-p_{10})\tilde{p},
\end{IEEEeqnarray}
where $\tilde{p}\triangleq p(A|\tilde{H}_{i,1},\tilde{H}_{i+1,1},\tilde{H}_{i+2,1})=p(s_i+n_i<0,s_{i+1}+n_{i+1}>0,s_{i+2}+n_{i+2}<0|\tilde{H}_{i,1},\tilde{H}_{i+1,1},\tilde{H}_{i+2,1})$. To compute $\tilde{p}$, the joint PDF $p(s_i+n_i,s_{i+1}+n_{i+1},s_{i+2}+n_{i+2}|\tilde{H}_{i,1},\tilde{H}_{i+1,1},\tilde{H}_{i+2,1})$ is needed. Here, we define $z_{i+2}\triangleq s_{i+2}+n_{i+2}$. Because $s_i$, $s_{i+1}$, and $s_{i+2}$ are jointly Gaussian, $z_i$, $z_{i+1}$, and $z_{i+2}$ will be also jointly Gaussian. So, the jointly PDF of $z_i$, $z_{i+1}$, and $z_{i+2}$ is
\begin{IEEEeqnarray}{rCl}
f(z_i,z_{i+1},z_{i+2})=\frac{1}{\sqrt{(2\pi)^3|C_3|}}\exp\{-\frac{1}{2}\zb^T_3C^{-1}_3\zb_3\},
\end{IEEEeqnarray}
where $\zb_3=[z_i, z_{i+1}, z_{i+2}]^T$, and $C_3$ is the covariance matrix of $\zb_3$. The elements of $C_3$ are $c_{3,11}=c_{3,22}=c_{3,33}=\sigma^2_1+\sigma^2$, $c_{3,12}=c_{3,21}=c_{3,23}=c_{3,32}=r\sigma_1^2$, and $c_{3,13}=c_{3,31}=0$. So, we have
\begin{IEEEeqnarray}{rCl}
\tilde{p}&=&\frac{1}{\sqrt{(2\pi)^3|C_3|}}\nonumber\\
&\times&\int_{-\infty}^{0}\int_{0}^{\infty}\int_{-\infty}^{0}\exp\{-\frac{1}{2}\zb^T_3C^{-1}_3\zb_3\}\mathrm{d}z_i\mathrm{d}z_{i+1}\mathrm{d}z_{i+2}.\nonumber\\
\end{IEEEeqnarray}
For the second case $j=i+2$, we have
\begin{IEEEeqnarray}{rCl}
\label{eq: pri14}
p^{'}_{00i(i+2)}&=&p(r_i\neq r_{i+1},r_{i+2}\neq r_{i+3}|H_1)\nonumber\\
&=&p(r_i=0,r_{i+1}=1,r_{i+2}=0,r_{i+3}=1|H_1)\nonumber\\
&+&p(r_i=0,r_{i+1}=1,r_{i+2}=1,r_{i+3}=0|H_1)\nonumber\\
&+&p(r_i=1,r_{i+1}=0,r_{i+2}=0,r_{i+3}=1|H_1)\nonumber\\
&+&p(r_i=1,r_{i+1}=0,r_{i+2}=1,r_{i+3}=0|H_1).\nonumber\\
\end{IEEEeqnarray}
We will compute $p(r_i=0,r_{i+1}=1,r_{i+2}=0,r_{i+3}=1|H_1)$ in (\ref{eq: pri14}). The other three probabilities can also be computed similarly. For $p(r_i=0,r_{i+1}=1,r_{i+2}=0,r_{i+3}=1|H_1)$, we have
\begin{IEEEeqnarray}{rCl}
p(&r_i&=0,r_{i+1}=1,r_{i+2}=0,r_{i+3}=1|H_1)\qquad\qquad\qquad\nonumber\\
&=&p(s_i+n_i<0,s_{i+1}+n_{i+1}>0,s_{i+2}+n_{i+2}<0\nonumber\\
&,&s_{i+3}+n_{i+3}>0).
\end{IEEEeqnarray}
If we define $C\triangleq s_i+n_i<0,s_{i+1}+n_{i+1}>0,s_{i+2}+n_{i+2}<0,s_{i+3}+n_{i+3}>0$, $p(r_i=0,r_{i+1}=1,r_{i+2}=0,r_{i+3}=1|H_1)$ can be written as
\begin{IEEEeqnarray}{rCl}
p(C)=p(\tilde{H}_{i,0})p(C|\tilde{H}_{i,0})+(1-p(\tilde{H}_{i,0}))p(C|\tilde{H}_{i,1}).
\end{IEEEeqnarray}
For $p(C|\tilde{H}_{i,0})$, we have
\begin{IEEEeqnarray}{rCl}
p(C|\tilde{H}_{i,0})=\frac{1}{2}p(D|\tilde{H}_{i,0}),
\end{IEEEeqnarray}
where $D\triangleq s_{i+1}+n_{i+1}>0,s_{i+2}+n_{i+2}<0,s_{i+3}+n_{i+3}>0$. For $p(D|\tilde{H}_{i,0})$, we have
\begin{IEEEeqnarray}{rCl}
p(D|\tilde{H}_{i,0})&=&p(\tilde{H}_{i+1,0}|\tilde{H}_{i,0})p(D|\tilde{H}_{i,0},\tilde{H}_{i+1,0})\nonumber\\
&+&p(\tilde{H}_{i+1,1}|\tilde{H}_{i,0})p(D|\tilde{H}_{i,0},\tilde{H}_{i+1,1})\nonumber\\
&=&(1-p_{01})p(D|\tilde{H}_{i,0},\tilde{H}_{i+1,0})\nonumber\\
&+&p_{01}p(D|\tilde{H}_{i,0},\tilde{H}_{i+1,1}).
\end{IEEEeqnarray}
For $p(D|\tilde{H}_{i,0},\tilde{H}_{i+1,0})$, we have
\begin{IEEEeqnarray}{rCl}
\label{eq: pri15}
p(D|\tilde{H}_{i,0},\tilde{H}_{i+1,0})&=&\frac{1}{2}p(E|\tilde{H}_{i+1,0})\nonumber\\
&=&p(\tilde{H}_{i+2,0}|\tilde{H}_{i+1,0})p(E|\tilde{H}_{i+1,0},\tilde{H}_{i+2,0})\nonumber\\
&+&p(\tilde{H}_{i+2,1}|\tilde{H}_{i+1,0})p(E|\tilde{H}_{i+1,0},\tilde{H}_{i+2,1})\nonumber\\
&=&\frac{1-p_{01}}{4}+p_{01}p(E|\tilde{H}_{i+1,0},\tilde{H}_{i+2,1}),\nonumber\\
\end{IEEEeqnarray}
where $E\triangleq s_{i+2}+n_{i+2}<0,s_{i+3}+n_{i+3}>0$. For computing $p(E|\tilde{H}_{i+1,0},\tilde{H}_{i+2,1})$, we can write as
\begin{IEEEeqnarray}{rCl}
\label{eq: pri17}
p(E|\tilde{H}_{i+1,0},\tilde{H}_{i+2,1})&=&p(E|\tilde{H}_{i+2,1})\nonumber\\
&=&p(\tilde{H}_{i+3,0}|\tilde{H}_{i+2,1})p(E|\tilde{H}_{i+2,1},\tilde{H}_{i+3,0})\nonumber\\
&+&p(\tilde{H}_{i+3,1}|\tilde{H}_{i+2,1})p(E|\tilde{H}_{i+2,1},\tilde{H}_{i+3,1})\nonumber\\
&=&\frac{p_{10}}{4}+(1-p_{10})(\frac{1}{2}-p).
\end{IEEEeqnarray}
For $p(D|\tilde{H}_{i,0},\tilde{H}_{i+1,1})$, we have
\begin{IEEEeqnarray}{rCl}
p(&D&|\tilde{H}_{i,0},\tilde{H}_{i+1,1})\nonumber\\
&=&p(\tilde{H}_{i+2,0}|\tilde{H}_{i+1,1})p(D|\tilde{H}_{i,0},\tilde{H}_{i+1,1},\tilde{H}_{i+2,0})\nonumber\\
&+&p(\tilde{H}_{i+2,1}|\tilde{H}_{i+1,1})p(D|\tilde{H}_{i,0},\tilde{H}_{i+1,1},\tilde{H}_{i+2,1})\nonumber\\
&=&\frac{p_{10}}{8}+(1-p_{10})p(D|\tilde{H}_{i,0},\tilde{H}_{i+1,1},\tilde{H}_{i+2,1}).
\end{IEEEeqnarray}
For computing $p(D|\tilde{H}_{i,0},\tilde{H}_{i+1,1},\tilde{H}_{i+2,1})$, we can write as
\begin{IEEEeqnarray}{rCl}
p(&D&|\tilde{H}_{i,0},\tilde{H}_{i+1,1},\tilde{H}_{i+2,1})=p(D|\tilde{H}_{i+1,1},\tilde{H}_{i+2,1})\nonumber\\
&=&p(\tilde{H}_{i+3,0}|\tilde{H}_{i+2,1})p(D|\tilde{H}_{i+1,1},\tilde{H}_{i+2,1},\tilde{H}_{i+3,0})\nonumber\\
&+&p(\tilde{H}_{i+3,1}|\tilde{H}_{i+2,1})p(D|\tilde{H}_{i+1,1},\tilde{H}_{i+2,1},\tilde{H}_{i+3,1})\nonumber\\
&=&\frac{p_{10}}{2}(\frac{1}{2}-p)+\tilde{p}^{'}(1-p_{10}),
\end{IEEEeqnarray}
where $\tilde{p}^{'}$ is defined as
\begin{IEEEeqnarray}{rCl}
\tilde{p}^{'}&\triangleq&\frac{1}{\sqrt{(2\pi)^3|C_3|}}\nonumber\\
&\times&\int_{0}^{\infty}\int_{-\infty}^{0}\int_{0}^{\infty}\exp\{-\frac{1}{2}\zb_3^TC^{-1}_3\zb_3\}\mathrm{d}z_i\mathrm{d}z_{i+1}\mathrm{d}z_{i+2}.\nonumber\\
\end{IEEEeqnarray}
For $p(C|\tilde{H}_{i,1})$, we have
\begin{IEEEeqnarray}{rCl}
p(C|\tilde{H}_{i,1})&=&p(\tilde{H}_{i+1,0}|\tilde{H}_{i,1})p(C|\tilde{H}_{i,1},\tilde{H}_{i+1,0})\nonumber\\
&+&p(\tilde{H}_{i+1,1}|\tilde{H}_{i,1})p(C|\tilde{H}_{i,1},\tilde{H}_{i+1,1})\nonumber\\
&=&p_{10}p(C|\tilde{H}_{i,1},\tilde{H}_{i+1,0})\nonumber\\
&+&(1-p_{10})p(C|\tilde{H}_{i,1},\tilde{H}_{i+1,1}).
\end{IEEEeqnarray}
For $p(C|\tilde{H}_{i,1},\tilde{H}_{i+1,0})$, we have
\begin{IEEEeqnarray}{rCl}
p(C|\tilde{H}_{i,1},\tilde{H}_{i+1,0})=\frac{1}{4}p(E|\tilde{H}_{i+1,0}),
\end{IEEEeqnarray}
where $p(E|\tilde{H}_{i+1,0})$ has been computed in (\ref{eq: pri15}), and (\ref{eq: pri17}). For $p(C|\tilde{H}_{i,1},\tilde{H}_{i+1,1})$, we have
\begin{IEEEeqnarray}{rCl}
p(&C&|\tilde{H}_{i,1},\tilde{H}_{i+1,1})\nonumber\\
&=&p(\tilde{H}_{i+2,0}|\tilde{H}_{i+1,1})p(C|\tilde{H}_{i,1},\tilde{H}_{i+1,1},\tilde{H}_{i+2,0})\nonumber\\
&+&p(\tilde{H}_{i+2,1}|\tilde{H}_{i+1,1})p(C|\tilde{H}_{i,1},\tilde{H}_{i+1,1},\tilde{H}_{i+2,1})\nonumber\\
&=&\frac{p_{10}}{4}(\frac{1}{2}-p)+(1-p_{10})p(C|\tilde{H}_{i,1},\tilde{H}_{i+1,1},\tilde{H}_{i+2,1}).\nonumber\\
\end{IEEEeqnarray}
For computing $p(C|\tilde{H}_{i,1},\tilde{H}_{i+1,1},\tilde{H}_{i+2,1})$, we can write as
\begin{IEEEeqnarray}{rCl}
p(&C&|\tilde{H}_{i,1},\tilde{H}_{i+1,1},\tilde{H}_{i+2,1})\nonumber\\
&=&p(\tilde{H}_{i+3,0}|\tilde{H}_{i+2,1})p(C|\tilde{H}_{i,1},\tilde{H}_{i+1,1},\tilde{H}_{i+2,1},\tilde{H}_{i+3,0})\nonumber\\
&+&p(\tilde{H}_{i+3,1}|\tilde{H}_{i+2,1})p(C|\tilde{H}_{i,1},\tilde{H}_{i+1,1},\tilde{H}_{i+2,1},\tilde{H}_{i+3,1})\nonumber\\
&=&\frac{p_{10}}{2}\tilde{p}+(1-p_{10})\bar{p},
\end{IEEEeqnarray}
where $\bar{p}\triangleq p(C|\tilde{H}_{i,1},\tilde{H}_{i+1,1},\tilde{H}_{i+2,1},\tilde{H}_{i+3,1})=p(s_i+n_i<0,s_{i+1}+n_{i+1}>0,s_{i+2}+n_{i+2}<0,s_{i+3}+n_{i+3}>0|\tilde{H}_{i,1},\tilde{H}_{i+1,1},\tilde{H}_{i+2,1},\tilde{H}_{i+3,1})$. To compute $\bar{p}$, the joint PDF $p(s_i+n_i,s_{i+1}+n_{i+1},s_{i+2}+n_{i+2},s_{i+3}+n_{i+3}|\tilde{H}_{i,1},\tilde{H}_{i+1,1},\tilde{H}_{i+2,1},\tilde{H}_{i+3,1})$ is needed. Now, we define $z_{i+3}\triangleq s_{i+3}+n_{i+3}$. Because $s_i$, $s_{i+1}$, $s_{i+2}$, and $s_{i+3}$ are jointly Gaussian, $z_i$, $z_{i+1}$, $z_{i+2}$, and $z_{i+3}$ will be also jointly Gaussian. So, the jointly PDF of $z_i$, $z_{i+1}$, $z_{i+2}$, and $z_{i+3}$ is
\begin{IEEEeqnarray}{rCl}
f(z_i,z_{i+1},z_{i+2},z_{i+3})=\frac{1}{\sqrt{(2\pi)^4|C_4|}}\exp\{-\frac{1}{2}\zb^T_4C^{-1}_4\zb_4\},\nonumber\\
\end{IEEEeqnarray}
where $\zb_4=[z_i, z_{i+1}, z_{i+2}, z_{i+3}]^T$, and $C_4$ is the covariance matrix of $\zb_4$. The elements of $C_4$ are $c_{4,11}=c_{4,22}=c_{4,33}=c_{4,44}=\sigma^2_1+\sigma^2$, $c_{4,12}=c_{4,21}=c_{4,23}=c_{4,32}=c_{4,34}=c_{4,43}=r\sigma_1^2$, and $c_{4,13}=c_{4,31}=c_{4,14}=c_{4,41}=c_{4,24}=c_{4,42}=0$. Therefore, we have
\begin{IEEEeqnarray}{rCl}
\bar{p}&=&\frac{1}{\sqrt{(2\pi)^4|C_4|}}\nonumber\\
&\times&\int_{0}^{\infty}\int_{-\infty}^{0}\int_{0}^{\infty}\int_{-\infty}^{0}\exp\{-\frac{1}{2}\zb^T_4C^{-1}_4\zb_4\}\mathrm{d}z_i\mathrm{d}z_{i+1}\mathrm{d}z_{i+2}\nonumber\\
&\qquad&\qquad\qquad\qquad\qquad\qquad\qquad\qquad\qquad\qquad\qquad\mathrm{d}z_{i+3}.\nonumber\\
\end{IEEEeqnarray}
For the third case $j\geq i+3$, we have
\begin{IEEEeqnarray}{rCl}
\label{eq: pri16}
p^{'}_{00ij}&=&p(r_i\neq r_{i+1},r_j\neq r_{j+1}|H_1)\nonumber\\
&=&p(r_i=0,r_{i+1}=1,r_j=0,r_{j+1}=1|H_1)\nonumber\\
&+&p(r_i=0,r_{i+1}=1,r_j=1,r_{j+1}=0|H_1)\nonumber\\
&+&p(r_i=1,r_{i+1}=0,r_j=0,r_{j+1}=1|H_1)\nonumber\\
&+&p(r_i=1,r_{i+1}=0,r_j=1,r_{j+1}=0|H_1).\nonumber\\
\end{IEEEeqnarray}
We will compute $p(r_i=0,r_{i+1}=1,r_j=0,r_{j+1}=1|H_1)$ in (\ref{eq: pri16}). The other three probabilities can also be computed similarly. For $p(r_i=0,r_{i+1}=1,r_j=0,r_{j+1}=1|H_1)$, we have
\begin{IEEEeqnarray}{rCl}
p(&r_i&=0,r_{i+1}=1,r_j=0,r_{j+1}=1|H_1)\qquad\qquad\qquad\nonumber\\
&=&p(s_i+n_i<0,s_{i+1}+n_{i+1}>0,s_j+n_j<0\nonumber\\
&,&s_{j+1}+n_{j+1}>0).
\end{IEEEeqnarray}
If we define $G\triangleq s_i+n_i<0,s_{i+1}+n_{i+1}>0,s_j+n_j<0,s_{j+1}+n_{j+1}>0$, $p(r_i=0,r_{i+1}=1,r_j=0,r_{j+1}=1|H_1)$ can be written as
\begin{IEEEeqnarray}{rCl}
p(G)=p(\tilde{H}_{i,0})p(G|\tilde{H}_{i,0})+(1-p(\tilde{H}_{i,0}))p(G|\tilde{H}_{i,1}).\nonumber\\
\end{IEEEeqnarray}
For $p(G|\tilde{H}_{i,0})$, we have
\begin{IEEEeqnarray}{rCl}
p(G|\tilde{H}_{i,0})=\frac{1}{2}p(I|\tilde{H}_{i,0}),
\end{IEEEeqnarray}
where $I\triangleq s_{i+1}+n_{i+1}>0,s_j+n_j<0,s_{j+1}+n_{j+1}>0$. For $p(I|\tilde{H}_{i,0})$, we have
\begin{IEEEeqnarray}{rCl}
p(I|\tilde{H}_{i,0})&=&p(\tilde{H}_{i+1,0}|\tilde{H}_{i,0})p(I|\tilde{H}_{i,0},\tilde{H}_{i+1,0})\nonumber\\
&+&p(\tilde{H}_{i+1,1}|\tilde{H}_{i,0})p(I|\tilde{H}_{i,0},\tilde{H}_{i+1,1})\nonumber\\
&=&(1-p_{01})p(I|\tilde{H}_{i,0},\tilde{H}_{i+1,0})\nonumber\\
&+&p_{01}p(I|\tilde{H}_{i,0},\tilde{H}_{i+1,1}).
\end{IEEEeqnarray}
For $p(I|\tilde{H}_{i,0},\tilde{H}_{i+1,0})$, we have
\begin{IEEEeqnarray}{rCl}
p(I|\tilde{H}_{i,0},\tilde{H}_{i+1,0})=p(I|\tilde{H}_{i+1,0})=\frac{1}{2}p(K|\tilde{H}_{i+1,0}),
\end{IEEEeqnarray}
where $K\triangleq s_j+n_j<0,s_{j+1}+n_{j+1}>0$. We can write $p(K|\tilde{H}_{i+1,0})$ as
\begin{IEEEeqnarray}{rCl}
\label{eq: pri18}
p(K|\tilde{H}_{i+1,0})&=&p(\tilde{H}_{j,0}|\tilde{H}_{i+1,0})p(K|\tilde{H}_{i+1,0},\tilde{H}_{j,0})\nonumber\\
&+&p(\tilde{H}_{j,1}|\tilde{H}_{i+1,0})p(K|\tilde{H}_{i+1,0},\tilde{H}_{j,1}),
\end{IEEEeqnarray}
where
\begin{IEEEeqnarray}{rCl}
[&p&(\tilde{H}_{j,0}|\tilde{H}_{i+1,0}), p(\tilde{H}_{j,1}|\tilde{H}_{i+1,0})]\nonumber\\
&=&[1, 0]\begin{bmatrix}
                                         1-p_{01} & p_{01} \\
                                         p_{10} & 1-p_{10}
                                       \end{bmatrix}^{j-i-1},
\end{IEEEeqnarray}
\begin{IEEEeqnarray}{rCl}
p(K|\tilde{H}_{i+1,0},\tilde{H}_{j,0})=p(K|\tilde{H}_{j,0})=\frac{1}{4},
\end{IEEEeqnarray}
and
\begin{IEEEeqnarray}{rCl}
\label{eq: pri19}
p(K|\tilde{H}_{i+1,0},\tilde{H}_{j,1})&=&p(K|\tilde{H}_{j,1})\nonumber\\
&=&p(\tilde{H}_{j+1,0}|\tilde{H}_{j,1})p(K|\tilde{H}_{j,1},\tilde{H}_{j+1,0})\nonumber\\
&+&p(\tilde{H}_{j+1,1}|\tilde{H}_{j,1})p(K|\tilde{H}_{j,1},\tilde{H}_{j+1,1})\nonumber\\
&=&\frac{p_{10}}{4}+(\frac{1}{2}-p)(1-p_{10}).
\end{IEEEeqnarray}
For $p(I|\tilde{H}_{i,0},\tilde{H}_{i+1,1})$, we have
\begin{IEEEeqnarray}{rCl}
p(I|\tilde{H}_{i,0},\tilde{H}_{i+1,1})&=&p(I|\tilde{H}_{i+1,1})\nonumber\\
&=&p(\tilde{H}_{j,0}|\tilde{H}_{i+1,1})p(I|\tilde{H}_{i+1,1},\tilde{H}_{j,0})\nonumber\\
&+&p(\tilde{H}_{j,1}|\tilde{H}_{i+1,1})p(I|\tilde{H}_{i+1,1},\tilde{H}_{j,1}),\nonumber\\
\end{IEEEeqnarray}
where
\begin{IEEEeqnarray}{rCl}
[&p&(\tilde{H}_{j,0}|\tilde{H}_{i+1,1}), p(\tilde{H}_{j,1}|\tilde{H}_{i+1,1})]\nonumber\\
&=&[0, 1]\begin{bmatrix}
                                         1-p_{01} & p_{01} \\
                                         p_{10} & 1-p_{10}
                                       \end{bmatrix}^{j-i-1},
\end{IEEEeqnarray}
\begin{IEEEeqnarray}{rCl}
p(I|\tilde{H}_{i+1,1},\tilde{H}_{j,0})=\frac{1}{8},
\end{IEEEeqnarray}
and
\begin{IEEEeqnarray}{rCl}
p(I|\tilde{H}_{i+1,1},\tilde{H}_{j,1})&=&p(\tilde{H}_{j+1,0}|\tilde{H}_{j,1})p(I|\tilde{H}_{i+1,1},\tilde{H}_{j,1},\tilde{H}_{j+1,0})\nonumber\\
&+&p(\tilde{H}_{j+1,1}|\tilde{H}_{j,1})p(I|\tilde{H}_{i+1,1},\tilde{H}_{j,1},\tilde{H}_{j+1,1})\nonumber\\
&=&\frac{p_{10}}{8}+\frac{1-p_{10}}{2}(\frac{1}{2}-p).
\end{IEEEeqnarray}
For $p(G|\tilde{H}_{i,1})$, we have
\begin{IEEEeqnarray}{rCl}
p(G|\tilde{H}_{i,1})&=&p(\tilde{H}_{i+1,0}|\tilde{H}_{i,1})p(G|\tilde{H}_{i,1},\tilde{H}_{i+1,0})\nonumber\\
&+&p(\tilde{H}_{i+1,1}|\tilde{H}_{i,1})p(G|\tilde{H}_{i,1},\tilde{H}_{i+1,1})\nonumber\\
&=&p_{10}p(G|\tilde{H}_{i,1},\tilde{H}_{i+1,0})\nonumber\\
&+&(1-p_{10})p(G|\tilde{H}_{i,1},\tilde{H}_{i+1,1}).
\end{IEEEeqnarray}
For $p(G|\tilde{H}_{i,1},\tilde{H}_{i+1,0})$, we have
\begin{IEEEeqnarray}{rCl}
p(G|\tilde{H}_{i,1},\tilde{H}_{i+1,0})=\frac{1}{4}p(K|\tilde{H}_{i,1},\tilde{H}_{i+1,0})=\frac{1}{4}p(K|\tilde{H}_{i+1,0}),\nonumber\\
\end{IEEEeqnarray}
where $p(K|\tilde{H}_{i+1,0})$ has been computed in (\ref{eq: pri18})-(\ref{eq: pri19}). For $p(G|\tilde{H}_{i,1},\tilde{H}_{i+1,1})$, we have
\begin{IEEEeqnarray}{rCl}
p(G|\tilde{H}_{i,1},\tilde{H}_{i+1,1})&=&p(\tilde{H}_{j,0}|\tilde{H}_{i+1,1})p(G|\tilde{H}_{i,1},\tilde{H}_{i+1,1},\tilde{H}_{j,0})\nonumber\\
&+&p(\tilde{H}_{j,1}|\tilde{H}_{i+1,1})p(G|\tilde{H}_{i,1},\tilde{H}_{i+1,1},\tilde{H}_{j,1}).\nonumber\\
\end{IEEEeqnarray}
For $p(G|\tilde{H}_{i,1},\tilde{H}_{i+1,1},\tilde{H}_{j,0})$, we have
\begin{IEEEeqnarray}{rCl}
p(&G&|\tilde{H}_{i,1},\tilde{H}_{i+1,1},\tilde{H}_{j,0})\nonumber\\
&=&p(\tilde{H}_{j+1,0}|\tilde{H}_{j,0})p(G|\tilde{H}_{i,1},\tilde{H}_{i+1,1},\tilde{H}_{j,0},\tilde{H}_{j+1,0})\nonumber\\
&+&p(\tilde{H}_{j+1,1}|\tilde{H}_{j,0})p(G|\tilde{H}_{i,1},\tilde{H}_{i+1,1},\tilde{H}_{j,0},\tilde{H}_{j+1,1})\nonumber\\
&=&\frac{1-p_{01}}{4}(\frac{1}{2}-p)+\frac{p_{01}}{4}(\frac{1}{2}-p)=\frac{1}{4}(\frac{1}{2}-p).
\end{IEEEeqnarray}
For $p(G|\tilde{H}_{i,1},\tilde{H}_{i+1,1},\tilde{H}_{j,1})$, we have
\begin{IEEEeqnarray}{rCl}
p(&G&|\tilde{H}_{i,1},\tilde{H}_{i+1,1},\tilde{H}_{j,1})\nonumber\\
&=&p(\tilde{H}_{j+1,0}|\tilde{H}_{j,1})p(G|\tilde{H}_{i,1},\tilde{H}_{i+1,1},\tilde{H}_{j,1},\tilde{H}_{j+1,0})\nonumber\\
&+&p(\tilde{H}_{j+1,1}|\tilde{H}_{j,1})p(G|\tilde{H}_{i,1},\tilde{H}_{i+1,1},\tilde{H}_{j,1},\tilde{H}_{j+1,1})\nonumber\\
&=&p_{10}p(G|\tilde{H}_{i,1},\tilde{H}_{i+1,1},\tilde{H}_{j,1},\tilde{H}_{j+1,0})\nonumber\\
&+&(1-p_{10})p(G|\tilde{H}_{i,1},\tilde{H}_{i+1,1},\tilde{H}_{j,1},\tilde{H}_{j+1,1}).
\end{IEEEeqnarray}
For $p(G|\tilde{H}_{i,1},\tilde{H}_{i+1,1},\tilde{H}_{j,1},\tilde{H}_{j+1,0})$, we have
\begin{IEEEeqnarray}{rCl}
p(G|\tilde{H}_{i,1},\tilde{H}_{i+1,1},\tilde{H}_{j,1},\tilde{H}_{j+1,0})&=&\frac{1}{2}p(M|\tilde{H}_{i,1},\tilde{H}_{i+1,1},\tilde{H}_{j,1})\nonumber\\
&=&\frac{1}{2}(\frac{1}{2}-p),
\end{IEEEeqnarray}
where $M\triangleq s_i+n_i<0,s_{i+1}+n_{i+1}>0,s_j+n_j<0$. For $p(G|\tilde{H}_{i,1},\tilde{H}_{i+1,1},\tilde{H}_{j,1},\tilde{H}_{j+1,1})$, we have
\begin{IEEEeqnarray}{rCl}
p(G|\tilde{H}_{i,1},\tilde{H}_{i+1,1},\tilde{H}_{j,1},\tilde{H}_{j+1,1})=(\frac{1}{2}-p)^2.
\end{IEEEeqnarray}


\begin{thebibliography}{1}

\bibitem{Kaydet98}
S. M.~Kay,
\newblock \emph{Fundamentals of statistical signal processing: detection theory},
\newblock Pearson, First edition, 1998.


\bibitem{Duarte06}
M. F. Duarte, M. A. Davenport, M. B. Wakin, and R. G. Baraniuk,
\newblock ``Sparse signal detection from incoherent projections,''
\newblock {\em In 2006 IEEE International Conference on Acoustics Speech and Signal Processing Proceedings}, France, 2006.

\bibitem{Haupt08}
J. Haupt, R. Castro, and R. Nowak,
\newblock ``Adaptive Discovery of Sparse Signals in Noise,''
\newblock {\em In 2008 42nd Asilomar Conference on Signals, Systems and Computers}, USA, 2008.

\bibitem{Wima17}
T. Wimalajeewa, and P. K. Varshney,
\newblock ``Sparse Signal Detection with Compressive Measurements via Partial Support Set Estimation,''
\newblock {\em IEEE Transactions on Signal and Information Processing over Networks}, vol. 3, No. 1, pp. 46--60, 2017.

\bibitem{Choi16}
J. Choi,
\newblock ``Sparse Signal Detection for Space Shift Keying using the Monte Carlo EM Algorithm,''
\newblock {\em IEEE Signal Processing Letters}, vol. 23, No. 7, pp. 974--978, 2016.

\bibitem{Naga18}
K. G. Nagananda, and P. K. Varshney,
\newblock ``On Weak Signal Detection With Compressive Measurements,''
\newblock {\em IEEE Signal Processing Letters}, vol. 25, No. 1, pp. 125--129, 2018.

\bibitem{Kafle18}
S. Kafle, T. Wimalajeewa, and P. K. Varshney,
\newblock ``Bayesian sparse signal detection exploiting laplace prior,''
\newblock {\em In 2018 IEEE International Conference on Acoustics, Speech and Signal Processing (ICASSP)}, Canada, 2018.

\bibitem{Wang18}
X. Wang, G. Li, and P. K. Varshney,
\newblock ``Detection of Sparse Signals in Sensor Networks via Locally Most Powerful Tests,''
\newblock {\em IEEE Signal Processing Letters}, vol. 25, No. 9, pp. 1418--1422, 2018.

\bibitem{Li20}
C. Li, G. Li, and P. K. Varshney,
\newblock ``Distributed Detection of Sparse Signals With Censoring Sensors Via Locally Most Powerful Test,''
\newblock {\em IEEE Signal Processing Letters}, vol. 27, pp. 346--350, 2020.

\bibitem{Mohammadi22}
A. Mohammadi, D. Ciuonzo, A. Khazaee, and P. Salvo Rossi,
\newblock ``Generalized Locally Most Powerful Tests for Distributed Sparse Signal Detection,''
\newblock {\em IEEE Transactions on Signal and Information Processing over Networks}, vol. 8, pp. 528--542, 2022.

\bibitem{Zuo21}
T. Zuo, F. Wang, and J.~Zhang,
\newblock ``Sparsity Signal Detection for Indoor GSSK-VLC System,''
\newblock {\em IEEE Trans. on Vehicular Technology.}, vol. 70, No. 12, pp. 12975--12984, 2021.

\bibitem{Han22}
R. Han, L. Bai, W. Zhang, J. Liu, J. Choi, and W.~Zhang,
\newblock ``Variational Inference Based Sparse Signal Detection for Next Generation Multiple Access,''
\newblock {\em IEEE Journal on selected areas in communications}, vol. 40, No. 4, pp. 1114--1127, 2022.

\bibitem{Li19}
C. Li, G. Li, B. Kailkhura, and P. K.~Varshney,
\newblock ``Secure Distributed Detection of Sparse Signals via Falsification of Local Compressive Measurements,''
\newblock {\em IEEE Trans. on Signal Proc.}, vol. 67, pp. 4696--4706, 2019.

\bibitem{Li20secure}
C. Li, G. Li, and P. K.~Varshney,
\newblock ``Distributed Detection of Sparse Signals With Physical Layer Secrecy Constraints: A Falsified Censoring Strategy,''
\newblock {\em IEEE Trans. on Signal Proc.}, vol. 68, pp. 6040--6054, 2020.

\bibitem{Wang19_1}
X. Wang, G. Li, and P. K.~Varshney,
\newblock ``Detection of Sparse Stochastic Signals With Quantized Measurements in Sensor Networks,''
\newblock {\em IEEE Trans. on Signal Proc.}, vol. 67, No. 8, pp. 2210--2220, 2019.

\bibitem{Wang19_2}
X. Wang, G. Li, C. Quan, and P. K.~Varshney,
\newblock ``Distributed Detection of Sparse Stochastic Signals With Quantized Measurements: The Generalized Gaussian Case,''
\newblock {\em IEEE Trans. on Signal Proc.}, vol. 67, No. 18, pp. 4886--4898, 2019.

\bibitem{Quan24}
C. Quan, Y. S. Han, B. Geng and P. K. Varshney,
\newblock ``Distributed Quantized Detection of Sparse Signals Under Byzantine Attacks,''
\newblock {\em IEEE Trans. Signal Process.}, vol. 72, No. 1, pp. 57--69, 2024.

\bibitem{Willett95}
P. Willett, and F.~Swaszek,
\newblock ``On the Performance Degradation from One-Bit Quantized Detection,''
\newblock {\em IEEE Trans. on Information Theory}, vol. 41, No. 6, pp. 1997--2003, 1995.

\bibitem{Fang13}
J. Fang, Y. Liu, H. Li, and S. Li,
\newblock ``One-Bit Quantizer Design for Multisensor GLRT Fusion,''
\newblock {\em IEEE Signal Processing Letters}, vol. 20, No. 3, pp. 257--260, 2013.

\bibitem{Zayy16}
H. Zayyani, F. Haddadi, and M. Korki,
\newblock ``Double detector for sparse signal detection from one-bit compressed sensing measurements,''
\newblock {\em IEEE Signal Processing Letters}, vol. 23, No. 11, pp. 1637--1641, 2016.

\bibitem{Wang19_3}
X. Wang, G. Li, and P. K.~Varshney,
\newblock ``Distributed Detection of Weak Signals From One-Bit Measurements Under Observation Model Uncertainties,''
\newblock {\em IEEE Signal Processing Letters}, vol. 26, No. 3, pp. 415--419, 2019.

\bibitem{Li19_1}
C. Li, Y. He, X. Wang, G. Li, and P. K.~Varshney,
\newblock ``Distributed Detection of Sparse Stochastic Signals via Fusion of 1-bit Local Likelihood Ratios,''
\newblock {\em IEEE Signal Processing Letters}, vol. 26, No. 12, pp. 1738--1742, 2019.

\bibitem{Li20tree}
C. Li, G. Li, and P. K.~Varshney,
\newblock ``Distributed Detection of Sparse Stochastic Signals With 1-Bit Data in Tree-Structured Sensor Networks,''
\newblock {\em IEEE Trans. on Signal Proc.}, vol. 68, pp. 2963--2976, 2020.

\bibitem{Ali18}
A. Ali, and W.~Hamouda,
\newblock ``Cooperative Low-Power Wideband Sensing Based on 1-bit Quantization,''
\newblock {\em IEEE Communications Letters}, vol. 22, No. 2, pp. 368--371, 2018.

\bibitem{Ali19}
A. Ali, and W.~Hamouda,
\newblock ``Generalized FFT-Based One-Bit Quantization System for Wideband Spectrum Sensing,''
\newblock {\em IEEE Trans. on Communications.}, vol. 68, No. 1, pp. 82--92, 2020.

\bibitem{Zayy20}
H. Zayyani, F. Haddadi, and M. Korki,
\newblock ``One-bit spectrum sensing in cognitive radio sensor networks,''
\newblock {\em Circuit, System, and Signal Processing (CSSP)}, vol. 39, pp. 2730--2743, 2020.

\bibitem{Zhao21}
Y. Zhao, X. Ke, B. Zhao, Y. Xiao, and L.~Huang,
\newblock ``One-Bit Spectrum Sensing Based on Statistical Covariances: Eigenvalue Moment Ratio Approach,''
\newblock {\em IEEE Wireless Communications Letters}, vol. 10, No. 11, pp. 2474--2478, 2021.

\bibitem{Ni23}
L. Ni, D. Zhang, Y. Sun, N. Liu, J. Liang, and Q.~Wan,
\newblock ``Detection and Localization of One-Bit Signal in Multiple Distributed Subarray Systems,''
\newblock {\em IEEE Trans. on Signal Proc.}, vol. 71, pp. 2776--2791, 2023.

\bibitem{Habibi21}
Z.~Habibi, and H. Zayyani, ``Markovian Adaptive Filtering Algorithm for
Block-Sparse System Identification,'' \emph{IEEE Trans. on Cir.
and Systems II: Exp. Briefs}, vol.~68, no.~8, pp. 3032--3036, August 2021.


%
%
%
%
%

%
%
%
%
%
%
%
%
%
%
%
%
%
%
%
%
%
%
%
%
%
%
%
%
%
%
%



\end{thebibliography}
\end{document}